\documentclass[copyright]{eptcs}
\providecommand{\event}{T. Neary, D. Woods, A.K. Seda and N. Murphy (Eds.):\\ The Complexity of Simple Programs 2008.}
\providecommand{\volume}{1}
\providecommand{\anno}{2009}
\providecommand{\firstpage}{31}
\usepackage{breakurl}

\usepackage{graphics}
\usepackage{latexsym}
\usepackage{amssymb}
\usepackage{amsfonts}

\newcommand{\ignore}[1]{}
\newcommand{\rst}{Rule 110}
\newcommand{\cts}{cyclic tag system}
\newcommand{\Cts}{Cyclic tag system}
\newcommand{\ts}{tag system}
\newcommand{\tm}{Turing machine}

\newcommand{\secref}[1]{section~\ref{#1}}

\newcommand{\figref}[1]{figure~\ref{#1}}

\newcommand{\pref}[1]{page~\pageref{#1}}

\newcommand{\segsec}[1]{\subsubsection*{\vspace{0pt}~~~\underline{#1}\vspace{0pt}}}
\newcommand{\mvleft}{\mbox{\it left}}
\newcommand{\mvright}{\mbox{\it right}}
\newcommand{\mvhalt}{\mbox{\it halt}}

\newcommand{\ebar}{$\bar{E}$\/}

\newcommand{\bb}[1]{$\mathbb{#1}$}

\newcommand{\figctr}[1]{\centerline{#1}} 

\newcommand{\mthctr}[1]{$#1$}

\newcommand{\up}[1]{$\nearrow _{\hspace{-1 ex} #1}$}

\newcommand{\upd}{$\nearrow$ distance}
\newcommand{\ov}[1]{$\stackrel{\raisebox{-1 ex}{$\scriptscriptstyle #1$}}{\frown}$}
\newcommand{\ovprot}[1]{$\stackrel{\protect\raisebox{-1 ex}{$\scriptscriptstyle #1$}}{\frown}$}

\newcommand{\ovd}{$\frown$ distance}

\newcommand{\smallcaption}[1]{\caption{#1}}

\newcommand{\updots}{\begin{picture}(0,0)\multiput(0,0)(-1,1){3}{\circle*{1}}\end{picture}}
\newcommand{\overdots}{\begin{picture}(0,0)\multiput(0,0)(0,1.5){3}{\circle*{1}}\end{picture}}
\newcommand{\squiggle}{\makebox(0,-9.5){\LARGE \~}}
\newcommand{\vsquiggle}{\makebox(0,0){\small $\wr$}}
\newcommand{\figdot}[3]{\put(#1,#2){\circle{8}}\put(#1,#2){\circle{2}}\put(#1,#2){\circle{5}}}
\newcommand{\sklab}[1]{\bf ({\em #1}\,)}
\newcommand{\grs}[2]{\sf \small #1}
\newcommand{\grAcc}{\makebox(0,0){\grs{acc}{ACC}}}
\newcommand{\grMd}{\makebox(0,2.5){\grs{md}{MD}}} 
\newcommand{\grPriStd}{\makebox(0,.5){\grs{pri/std}{PRI/STD}}}
\newcommand{\grStd}{\makebox(0,2.5){\grs{std}{STD}}}
\newcommand{\grRej}{\makebox(0,.5){\grs{rej}{REJ}}}
\newcommand{\grTd}{\makebox(0,2.5){\grs{td}{TD}}}
\newcommand{\grInv}{\makebox(0,2.5){\grs{inv}{INV}}}
\newcommand{\grPrep}{\makebox(0,-1.5){\grs{prep}{PREP}}}
\newcommand{\grPri}{\makebox(0,.5){\grs{pri}{PRI}}}
\newcommand{\grAccRej}{\makebox(0,.5){\grs{acc/rej}{ACC/REJ}}}
\newcommand{\grInvMd}{\makebox(0,.5){\grs{inv/md}{INV/MD}}}
\newcommand{\grRaw}{\makebox(0,0){\grs{raw}{RAW}}}
\newcommand{\grShortPrep}{\makebox(0,7.5){$\stackrel{\mbox{\sf \small short}}{\mbox{\sf \small prep}}$}}
\newcommand{\beginpicture}[2]{\begin{picture}(#1,#2)
}
\newcommand{\eendpicture}{\end{picture}} 

	\title{A Concrete View of Rule 110 Computation}
	\author{Matthew Cook
        \email{}
	    \institute{Institute of Neuroinformatics, University of Zurich and ETH Zurich}
        }

\def\titlerunning{A Concrete View of Rule 110 Computation}
\def\authorrunning{M. Cook}

\begin{document}
\maketitle

	\begin{abstract}
		\rst\ is a cellular automaton that performs repeated simultaneous updates of an
		infinite row of binary values.  The values are updated in the following way:
		0s are changed to 1s at all positions where the value to the right is a~1,
		while 1s are changed to 0s at all positions where the values to the left
		and right are both~1.
		Though trivial to define, the behavior exhibited by \rst\ is
		surprisingly intricate, and in \cite{Cook04a} we showed that it is capable
		of emulating the activity of a \tm\ by encoding the \tm\ and its tape
		into a repeating left pattern, a central pattern, and a repeating right pattern,
		which \rst\ then acts on.
		In this paper we provide an explicit compiler for converting a Turing machine into a \rst\ initial state,
		and we present a general approach for proving that such constructions will work as intended.
		The simulation was originally assumed to require exponential time,
		but surprising results of Neary and Woods \cite{Neary06} have shown
		that in fact, only polynomial time is required.
		We use the methods of Neary and Woods to exhibit
		a direct simulation of a \tm\ by a \ts\ in polynomial time.
	\end{abstract}

\section{Compiling a \tm\ into a \rst\ State}\label{compiler}

In this section we give a concrete algorithm for compiling
a \tm\ and its tape into an initial state for \rst , following
the construction given in \cite{Cook04a}.
We will create an initial state that
will eventually produce the bit sequence $01101001101000$
if and only if the corresponding \tm\ halts.
Or, if one prefers time sequences to spatial sequences,
    it is also the case that the sequence
    $110101010111111$ will be produced over time by a single cell
    if and only if the \tm\ halts.
    (These sequences are the shortest that occur for the first time
    in the collision that produces the $F$ glider
    as in \figref{figSketchesZ}(z), which only occurs if the algorithm halts.)
While based directly on the methods of \cite{Cook04a},
the presentation of the algorithm here is self-contained.
This section can be viewed as heavily commented high level pseudocode,
the intent being to explicitly provide all the details a program would use,
regardless of programming language or input or output format.

Following \cite{Cook04a},
we will convert the \tm\ into a \ts , which we then convert into
a \cts , and finally into a bit sequence for \rst .
%
%
Of course, if one is starting with a \ts ,
 or a \cts\ (whose appendants' lengths are multiples of 6),
 then the unnecessary
conversions may be omitted.
  For example, Paul Chapman's \ts\ \cite{Chapman03}
  for the $3x+1$ problem, given by
  $\{A \leadsto C,  B \leadsto D,  C \leadsto A E,
     D \leadsto B F,  E \leadsto C C D,  F \leadsto D D D \}$\label{ChapmansTag},
  with a starting tape of $C[D]^{x-1}$,
  or Liesbeth De Mol's more recent \ts\ for the same problem \cite{DeMol08},
  $\{A \leadsto CY, C \leadsto A, Y \leadsto AAA\}$,
  can be implemented nicely in \rst\ without going through a \tm\ representation.
  As another example, the \cts\ $\{YYYYYY,\emptyset,NNNNNN,\emptyset\}$,
  starting with a single $Y$ on the tape,
  yields exactly the same behavior as the \cts\ shown in figure~2 of~\cite{Cook04a}
  with runs of $Y$s (after the initial $Y$)
  doubled and runs of $N$s lengthened by a factor of six.
  This \cts\ is also shown on
    Page 96 of~\cite{Wolfram02}, which
  shows a random-looking graph of its behavior
  during the first million steps.

If we start with a \cts\ whose appendants are not each a multiple of six long,
we can convert it into such a form as follows:
Expand each appendant by adding 5 $N$ symbols after every symbol,
and expand the list of appendants by adding 5 empty appendants after every appendant.
The tape should also be expanded just as each appendant was.
This will make every appendant's length be a multiple of six,
while performing the same computation, on every sixth step, as the original \cts .

\subsection{We start with a Turing machine}

Suppose we are given a \tm\ with $m$ states
\mthctr{\Psi = \{\psi_1, \psi_2, \ldots , \psi_m\}}
and $t$ symbols\linebreak \mthctr{\Sigma = \{\sigma_1, \sigma_2, \ldots, \sigma_t\}}
and we are given its lookup tables for
which symbol to write \mthctr{\Upsilon(\psi_i,\sigma_j) \in \Sigma},
which way to move \mthctr{\Delta(\psi_i,\sigma_j) \in \{\mvleft,\mvright,\mvhalt\}},
and what state to go into \mthctr{\Gamma(\psi_i,\sigma_j) \in \Psi}.
($\Upsilon$ and $\Gamma$ do not need to be defined for (state, symbol)
pairs that cause the machine to halt, i.e.\ for which $\Delta$ is $\mvhalt$.)

Suppose further that the \tm\ is currently in state $\psi_{\gamma}$,
and that the tape is of the form
\[
\overline{\sigma_{a_w} \: \sigma_{a_{w-1}} \cdots \: \sigma_{a_1}}
\: \sigma_{b_x} \: \sigma_{b_{x-1}} \cdots \: \sigma_{b_1}
\: \fbox{$\sigma_c$}
\: \sigma_{d_1} \: \sigma_{d_2} \cdots \: \sigma_{d_y}
\: \overline{\sigma_{e_1} \: \sigma_{e_2} \cdots \: \sigma_{e_z}},\]
where the box represents the current position of the \tm 's head,
and the portions with lines over them are repeated into the distance
of the two-way infinite tape.

Our task is to convert this complete description of the \tm\ and its infinite tape
into an initial state for \rst .

\subsection{We transform it into a tag system}

Our first transformation will be to convert this \tm\ into a set
of \ts\ rules for a \ts\ with deletion number $s$, meaning that
at each step, $s$ symbols are removed from the front of the tape,
and an appendant is appended according only to the first of the $s$ removed symbols.
There are two general approaches for doing this.
The traditional (since 1964) approach of Cocke and Minsky~\cite{Cocke64}
is simpler, but results in an exponential slowdown in simulation time.
The modern (since 2006) approach of Neary and Woods~\cite{Neary06}
is more complicated, but amazingly solves what I call the
``geometry problem'' of cyclic-state tape processors
such as tag systems and cyclic tag systems.
The geometry\label{geomprob} problem in this instance is that the processing head
is unaware, as it scans the tape, of which tape symbols are next to which other tape symbols.
This is
because it effectively has a fixed number of bits of memory,
consisting of the phase of the head with respect to the tape,
and these bits can only be read or written with difficulty.
One difficulty is that the bits must be shared for an entire pass of processing the tape,
meaning that the machine is unable to simply remember at each step what symbol it just saw,
as a Turing machine might do.
Indeed, everything the machine does remember is due to
the sum total effect of the previous pass, rather than the current pass.
This makes it nearly impossible for the system
to know the ordering of the symbols on the tape.
Without being able to detect the order of the symbols,
encodings are limited to a unary representation,
which makes the tape exponentially larger than a binary representation,
thus making passes over the tape take exponentially longer.

In this section we will perform the transformation using the
simpler method of Cocke and Minsky.
For an explanation of why it is that a transformation like this
results in a \ts\ that correctly emulates the \tm , see~\cite{Cook04a}.
Here we just focus on the mechanics of the transformation itself.
In \secref{directsim} we will show how
a conversion like this can be done using the more
complicated method of Neary and Woods.

To enable our transformation, we will first add two symbols to the alphabet,
$\sigma_{t+1}$ and $\sigma_{t+2}$, for a total of $s=t+2$ symbols.
These two new symbols will be used to
mark the left and right ends of
the central nonperiodic portion of the tape.
We will not need to define $\Upsilon$, $\Delta$, or $\Gamma$ for
$\sigma_{t+1}$ or $\sigma_{t+2}$.

Now we can transform the \tm\ into a set of \ts\ rules on
an alphabet $\Phi$ of $4m+3ms$ symbols:
\[
\Phi = \left \{ \!\!\! \begin{array}{cl}
   H_{\psi_i}, L_{\psi_i}, R_{\psi_i}, R_{\psi_i\ast} &
     \;\; \mbox{for each $i \in \{1\ldots m\}$}
\\[1ex]
   H_{\psi_i \sigma_j}, L_{\psi_i \sigma_j}, R_{\psi_i \sigma_j} &
     \;\; \mbox{for each $i \in \{1\ldots m\}$ and $j \in \{1\ldots s\}$}
\end{array} \!\!\! \right \}\]
The \ts\ rules, one for each symbol in $\Phi$,
are based on $\Upsilon$, $\Delta$, and $\Gamma$
as follows:

\begin{center}
\mthctr{
\begin{array}{c}
 \begin{array}{rcl}
  H_{\psi_i} & \leadsto & H_{\psi_i \sigma_1} H_{\psi_i \sigma_2} \ldots H_{\psi_i \sigma_{s}}
 \\[1 ex]
  L_{\psi_i} & \leadsto & L_{\psi_i \sigma_1} L_{\psi_i \sigma_2} \ldots L_{\psi_i \sigma_{s}}
 \\[1 ex]
  R_{\psi_i} & \leadsto & R_{\psi_i \sigma_1} R_{\psi_i \sigma_2} \ldots R_{\psi_i \sigma_{s}}
 \\[1 ex]
  R_{\psi_i\ast} & \leadsto & [R_{\psi_i}]^{s}
 \\[1 ex]
  H_{\psi_i \sigma_j | \Delta(\psi_i, \sigma_j)=\mvleft} & \leadsto &
    [R_{\Gamma(\psi_i, \sigma_j)\ast}]^{s (s - \Upsilon(\psi_i, \sigma_j))}
    [H_{\Gamma(\psi_i, \sigma_j)}]^{j}
 \\[1 ex]
  H_{\psi_i \sigma_j | \Delta(\psi_i, \sigma_j)=\mvright} & \leadsto &
    [H_{\Gamma(\psi_i, \sigma_j)}]^{j}
    [L_{\Gamma(\psi_i, \sigma_j)}]^{s (s - \Upsilon(\psi_i, \sigma_j))}
 \\[1 ex]
  H_{\psi_i \sigma_j | \Delta(\psi_i, \sigma_j)=\mvhalt} & \leadsto & \emptyset
 \\[1 ex]
  H_{\psi_i \sigma_{t+1}} & \leadsto & [H_{\psi_i}]^{t+1+s-a_1}
        [L_{\psi_i}]^{s^w + \sum_{k=2}^{w} (s-a_k) s^{k-1}}
 \\[1 ex]
  H_{\psi_i \sigma_{t+2}} & \leadsto & [R_{\psi_i \ast}]^{\sum_{k=2}^{z} (s-e_k) s^{k-1}}
        [H_{\psi_i}]^{t+2+s-e_1}
 \end{array}
\end{array}
}
\end{center}
\begin{center}
\mthctr{
\begin{array}{c}
 \begin{array}{rcl}
  L_{\psi_i \sigma_j | \Delta(\psi_i, \sigma_j)=\mvleft} & \leadsto & L_{\Gamma(\psi_i, \sigma_j)} 
 \\[1 ex]
  L_{\psi_i \sigma_j | \Delta(\psi_i, \sigma_j)=\mvright} & \leadsto & [L_{\Gamma(\psi_i, \sigma_j)}]^{s^2}
 \\[1 ex]
  L_{\psi_i \sigma_j | \Delta(\psi_i, \sigma_j)=\mvhalt} & \leadsto & \emptyset
 \\[1 ex]
  L_{\psi_i \sigma_{t+1 \:\mbox{\tiny or } t+2}} & \leadsto & [L_{\psi_i}]^{s}
 \\[1 ex]
  R_{\psi_i \sigma_j | \Delta(\psi_i, \sigma_j)=\mvleft} & \leadsto & [R_{\Gamma(\psi_i, \sigma_j)}]^{s^2}
 \\[1 ex]
  R_{\psi_i \sigma_j | \Delta(\psi_i, \sigma_j)=\mvright} & \leadsto & R_{\Gamma(\psi_i, \sigma_j)}
 \\[1 ex]
  R_{\psi_i \sigma_j | \Delta(\psi_i, \sigma_j)=\mvhalt} & \leadsto & \emptyset
 \\[1 ex]
  R_{\psi_i \sigma_{t+1 \:\mbox{\tiny or } t+2}} & \leadsto & [R_{\psi_i}]^{s}
 \end{array}
\end{array}
}
\end{center}
The notation $[symbol]^{n}$ represents $n$ consecutive copies of the symbol.
Each symbol whose subscript is a
pair $(\psi_i, \sigma_j)$ that causes the \tm\ to halt
leads to an empty right hand side,
denoted by $\emptyset$.
If the \tm\ does not halt, these empty right hand sides will never be used.

Additionally, we transform the \tm\ tape into the following \ts\ tape:
\[
[ H_{\psi_{\gamma}} ]^{1+s-c} \:\:
[ L_{\psi_{\gamma}} ]^{s^{x+1} + \sum_{k=1}^{x} (s-b_k) s^{k}} \:
[ R_{\psi_{\gamma}} ]^{\sum_{k=1}^{y} (s-d_k) s^{k}}
\]

This completes our transformation of the \tm\ into a \ts\ with deletion
number $s$.

  If for some reason one wants to to avoid \ts\ rules
  that are exponentially long (in the length of the \tm 's tape's periodicity),
  then instead of using one new symbol at the end of the tape to extend the
  tape by another period, one can use many new symbols, each of which extends
  the tape by a limited amount.
  Similarly, if one wants to avoid an initial \ts\ tape that is exponentially
  long (in the length of the initial \tm\ tape),
  new states can be added to the \tm\ for the sole purpose of writing the
  initial tape and positioning the \tm\ on it.
  Using these two methods, the entire compilation algorithm in this section takes only
  polynomial time in the size of the \tm 's initial
  configuration, and creates a \rst\ initial
  state of polynomial size.

\subsection{We further transform it into a cyclic tag system}

Our next transformation will be to convert the \ts\ tape and rules
into a tape and cyclic appendant sequence for a \cts .

\Cts s were invented as part of the proof
of \rst 's universality,
but they are also interesting in their own right.
For example, Neary and Woods
introduced their method in~\cite{Neary06} by showing how a \cts\ can emulate a \tm .
They simply cycle through a list of appendants as they read the tape,
appending the current appendant to the end of the tape whenever a $Y$ is read.

To begin this transformation, we assign an ordering to the \ts\ alphabet $\Phi$:
\mthctr{
\phi_{1}=H_{\psi_{1}},\:\:
\phi_{2}=H_{\psi_{2}},\:\:
\ldots,\:\: \phi_{4m+3ms}=R_{\psi_{m}\sigma_{s}}}
Next, we extend $\Phi$
so that its size $|\Phi|$ becomes a multiple of $6$,
by adding anywhere from 0 to 5 dummy rules to the \ts , of the form:
\mthctr{
\phi_{4m+3ms+1} \leadsto \emptyset, \:\:\:
\phi_{4m+3ms+2} \leadsto \emptyset, \:\:\:
\ldots}

Now we can create the \cts\ by listing the \ts 's rules in order,
and converting each right hand side into a string of $Y$s and $N$s
via a simple unary encoding where each $\phi_{i}$ becomes
a string of $|\Phi|$ $N$s with the $i^{th}$ one changed to a $Y$:
\mthctr{\phi_{i}\:\rightarrowtail\:[N]^{i-1} \: Y \: [N]^{|\Phi|-i}}

So for example the first rule, \mthctr{
  H_{\psi_1} \:\: \leadsto \:\:
   H_{\psi_1 \sigma_1} H_{\psi_1 \sigma_2} \ldots H_{\psi_1 \sigma_{s}}
} can be rewritten using the $\phi_{i}$s 
 as \mthctr{
  \phi_{1} \:\: \leadsto \:\:
   \phi_{4m+1} \phi_{4m+2} \ldots \phi_{4m+s}
} whose right hand side gets converted to \linebreak
\mthctr{
  N^{4m} \: Y \: N^{|\Phi|-4m-1} \: N^{4m+1} \: Y \: N^{|\Phi|-4m-2} \:
        \ldots \: N^{4m+s-1} \: Y \: N^{|\Phi|-4m-s}
} which happens to simplify to\linebreak \mthctr{
  N^{4m} \: Y \: [ N^{|\Phi|} \: Y ] ^{s-1} \: N^{|\Phi|-4m-s}
} and so this is the first appendant in the \cts 's cyclic list.

The \cts 's cyclic list starts with the $|\Phi|$ appendants that can
be generated in this way from the rules of the \ts ,
and then it is extended to a length of $s |\Phi|$ by
simply adding $(s-1) |\Phi|$ empty appendants.

The initial tape for the \cts\ is simply the unary encoding of
the \ts 's tape into $Y$s and $N$s.  This completes our conversion
of the \ts\ into a \cts\ with $s |\Phi|$ appendants.

\subsection{We finally convert it into a Rule 110 state}

Our final transformation will be to convert the \cts 's tape and appendants into
an initial state for \rst .
We will do this by simply gluing together bit sequences for
the various glider clusters involved.
We will start with the central tape region, and then we will
specify the periodic sequences to its right and left.

\begin{figure}
 \figctr{
   $\stackrel{\mathbb{A}}{\mbox{\includegraphics{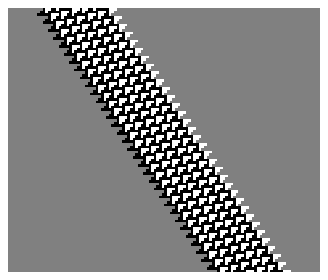}}}$
   $\stackrel{\mathbb{B}}{\mbox{\includegraphics{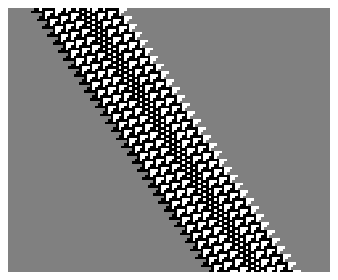}}}$
 }
 \figctr{
   $\stackrel{\mathbb{C}}{\mbox{\includegraphics{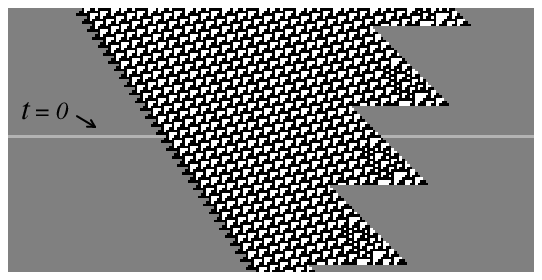}}}$
   $\stackrel{\mathbb{D}}{\mbox{\includegraphics{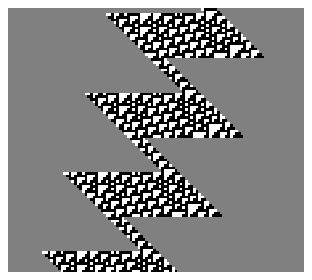}}}$
 }
 \figctr{
   $\stackrel{\mathbb{E}}{\mbox{\includegraphics{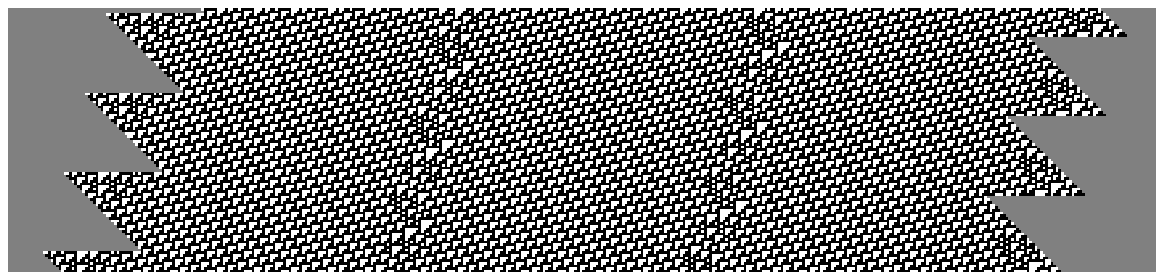}}}$
 }
 \figctr{
   $\stackrel{\mathbb{F}}{\mbox{\includegraphics{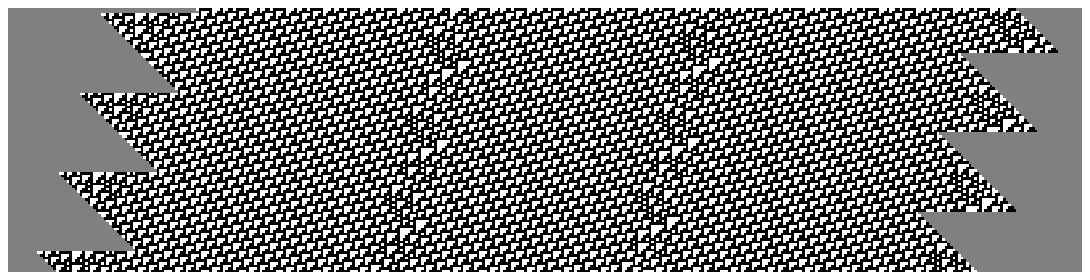}}}$
 }
 \figctr{
   $\stackrel{\mathbb{G}}{\mbox{\includegraphics{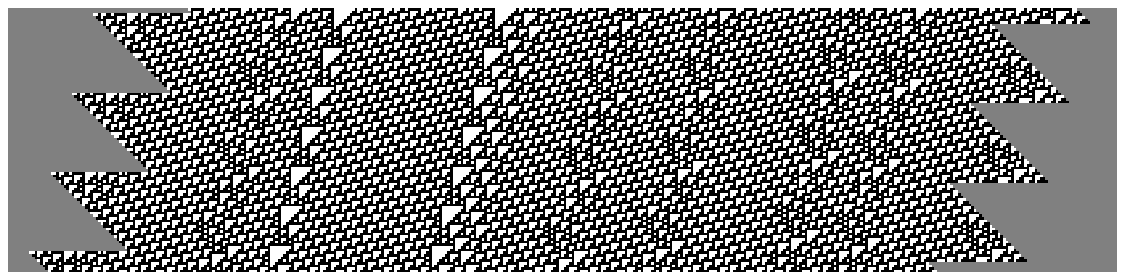}}}$
 }
 \smallcaption{
  \label{figBitBlocksLC}
  Blocks of bits used in generating the initial state for \rst .
  Two blocks are joined by simply fitting them together along the
  zig-zag edge so that the $t=0$ row,
  defined in block \bb{C},
  gets extended into the new block.
  Each block except for \bb{C} is periodic:
  Blocks \bb{A} and \bb{B} repeat every $3$ lines, and
  the other blocks repeat every $30$ lines.
  The cyclic left hand side of \rst 's initial state
  is built with blocks \bb{A} and \bb{B}.
  The central region is built with blocks \bb{C} - \bb{G}.
 }
\end{figure}

\begin{figure}
 \figctr{
   $\stackrel{\mathbb{H}}{\mbox{\includegraphics{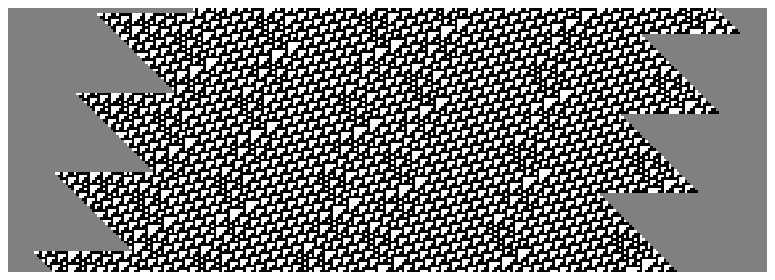}}}$
 }
 \figctr{
   $\stackrel{\mathbb{I}}{\mbox{\includegraphics{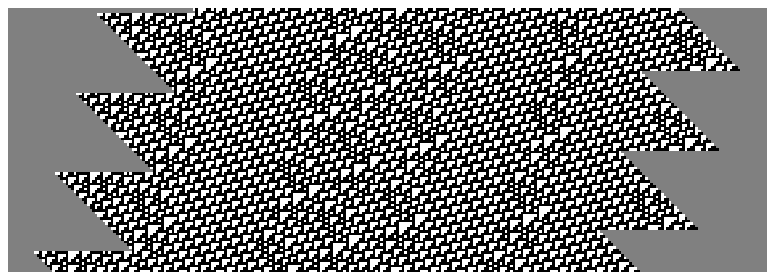}}}$
 }
 \figctr{
   $\stackrel{\mathbb{J}}{\mbox{\includegraphics{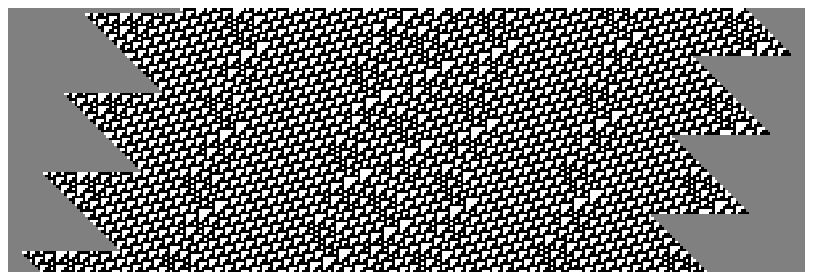}}}$
 }
 \figctr{
   $\stackrel{\mathbb{K}}{\mbox{\includegraphics{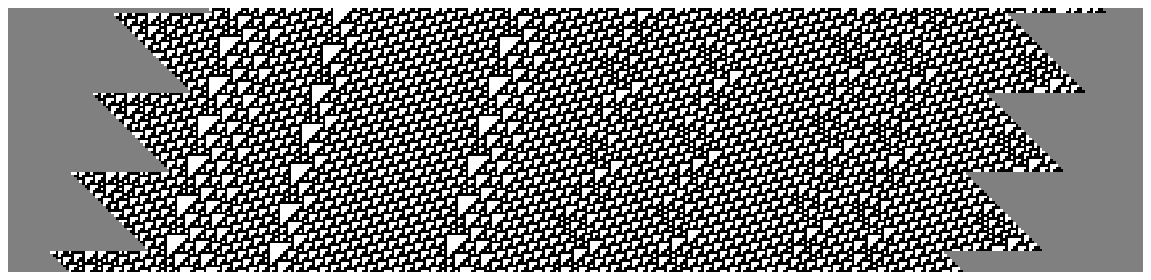}}}$
 }
 \figctr{
   $\stackrel{\mathbb{L}}{\mbox{\includegraphics{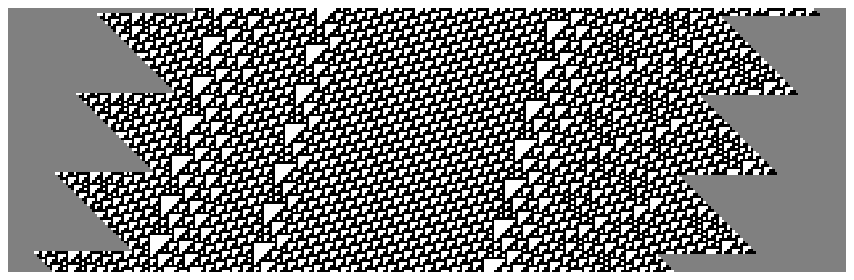}}}$
 }
 \smallcaption{
  \label{figBitBlocksR}
  More blocks of bits used in generating the initial state for \rst .
  The blocks in this figure are used to create
  the cyclic right hand side of \rst 's initial state
  based on the \cts 's appendant list.
  A program for creating the \rst\ initial state
  would just store 30 strings of bits for each block,
  along with the vertical phase offset
  of the right hand zig-zag from the left hand one.
  Note that there is an \ebar\ glider traveling right in the middle
  of every zig-zag region, so two blocks always join at an \ebar .
 }
\end{figure}

We will start the central bit sequence with the row marked
in \figref{figBitBlocksLC}, block \bb{C}\/.
We will first extend this row to the right by attaching other blocks.
Each time we attach another block, we do it so that the zig-zag
seam fits together perfectly,
as if the two blocks were pieces of a large simple jigsaw puzzle,
without worrying about whether the top or bottom edges of the blocks are aligned.
Usually, the new block will not be able to be at exactly
the same height as the previous block,
but we simply need to make sure that the new block extends
the $t=0$ row marked in block \bb{C}\/.
This row will be the initial
state for \rst .
Programmatically, ``attaching the blocks'' is a
very simple process:  As we extend the initial state,
we just keep track of which phase of the zig-zag the $t=0$ row is
at every time we cross from one block to another.  This phase
tells us which row to use from the new block, and then the phase
of the zig-zag at the end of that same row in the new block
becomes the phase for the next crossing.

The central region is formed fairly directly from the \cts 's tape:
Each $N$ becomes \bb{ED}, and each $Y$ becomes \bb{FD}, but then the
very last \bb{D} is changed to \bb{G}, and \bb{C} is stuck on the front.
So for example, a tape of $NNYN$ would become \bb{CEDEDFDEG}\/.

Next, we form the periodic right hand side with a periodic
sequence of blocks based on the \cts 's appendants.
Each appendant from the \cts 's list is converted by changing
each $Y$ to \bb{II} and each $N$ to \bb{IJ}, but then the very
first \bb{I} is replaced with \bb{KH}.  If there is no first \bb{I},
due to the appendant being empty, then an \bb{L} is used for that
appendant.  Once this is done for all appendants, the initial \bb{K}
of the first appendant is moved to the very end.
So for example, the cyclic appendant list
$\{YN,\:NYYN,\:\emptyset,\:\emptyset\}$ would become \bb{HIIJKHJIIIIIJLLK}\/.
This sequence of blocks gets repeated on the right hand side,
and so the bit sequence for the $t=0$ row must also be periodic,
since after some number of repetitions of the sequence of blocks,
the bit sequence will enter the same row of the initial block as it
did at the very beginning, and so the bit sequence becomes periodic
at that point.

The periodic left side has the form:
\mthctr{
    [\mathbb{A}]^{v}\:\mathbb{B}\:
    [\mathbb{A}]^{13}\:\mathbb{B}\:
    [\mathbb{A}]^{11}\:\mathbb{B}\:
    [\mathbb{A}]^{12}\:\mathbb{B}
}

\mthctr{
   \begin{array}{rcl}
    \mbox{where } v =
    &   & 76 \cdot \mbox{(the total number of $Y$s in all appendants)} \\
    & + & 80 \cdot \mbox{(the total number of $N$s in all appendants)} \\
    & + & 60 \cdot \mbox{(the number of nonempty appendants)} \\
    & + & 43 \cdot \mbox{(the number of empty appendants)}
    \end{array}
}

Calculating the periodic sequence of bits for the $t=0$ row
on the left works just
like calculating the sequence on the right, except that we work
our way to the left from the $t=0$ row in block \bb{C}, crossing
over from block \bb{C} to block \bb{B}, then twelve copies of
block \bb{A}, and so on.  The bit sequence will have to go through
the $[\mathbb{A}]^{v}\mathbb{B}\:[\mathbb{A}]^{13}\mathbb{B}\:
    [\mathbb{A}]^{11}\mathbb{B}\:[\mathbb{A}]^{12}\mathbb{B}$
block sequence three times before it starts repeating.

This completes our algorithm for transforming an arbitrary \tm\ into
an initial state for \rst\ consisting of a periodic sequence of bits
on the left, followed by a central sequence, followed by a periodic
sequence on the right.

\subsection{Some comments on this algorithm}

The blocks in figures~\ref{figBitBlocksLC} and \ref{figBitBlocksR}
correspond to the conceptual clusters of gliders used in the construction in~\cite{Cook04a}.

For example, the $\mathbb{B}\:
    \mathbb{A}^{13}\:\mathbb{B}\:
    \mathbb{A}^{11}\:\mathbb{B}\:
    \mathbb{A}^{12}\:\mathbb{B}
$ assembly is an {\it ossifier}, containing an $A^4$ in each \bb{B},
separated by pure {\it ether} in the \bb{A}s.

The \bb{E} block, including the \ebar s at its seams,
 is a {\it moving data N},
  while the \bb{F} block is a {\it moving data Y}\@.
The \bb{D} block glues adjacent elements of moving data together.

The remaining blocks each contain a glider cluster,
extending from the first fully present glider to the \ebar\ in the right seam,
so only the \ebar\ in the left seam is not a member of the cluster.
The \ebar\ in the left seam is simply the last \ebar\ of the previous cluster,
thus providing the position of the block's cluster in relation to the previous cluster.

The \bb{H} is a {\it primary component},
while \bb{I} and \bb{J} are {\it standard components},
differing only in the spacing from the previous component.
The \bb{J} uses more space,
so that \bb{IJ} encodes an $N$ of {\it table data},
while \bb{II} encodes a $Y$ of {\it table data}.

The \bb{K} block is a {\it raw leader}, and the \bb{L} block is a {\it raw short leader}.
The \bb{G} block is a {\it prepared leader}.
There is no block shown for a {\it prepared short leader},
since our algorithm assumed that the first appendant would not be empty.
If the first appendant is empty, we would need a block for a prepared short leader,
to be used in place of the \bb{G}.
This block would be a modified form of \bb{L}
using exactly the same modification
that turned the \bb{K} into the \bb{G}.

The calculation of $v$ for the periodic left hand side
corresponds to a conservatively large rough estimate
of twice the total vertical height of all the table data, including both components and leaders.
This is used to set the vertical spacing between ossifiers
so that ossifiers do not hit tape data by mistake.
This assumes that at least one nonempty appendant will be appended on each
cycle through the appendants if the system is not halting.
This will indeed be the case if the system has been compiled from a \tm\ as described
above, since the transformation to a \ts\ specifically ensures this property.
If the system was compiled directly from a \ts\ or \cts ,
then an appropriate value of $v$ can be chosen if a bound can
be placed on the number of consecutively rejected or empty appendants.
If there is no such bound, then it will not be possible for this construction
to work with a periodic left hand side.  In this case, a more complicated left hand side
will work, where the value of $v$ increases linearly with each ossifier,
since the length of the tape can only increase linearly with time,
and clearly at any time the length of the tape constitutes a bound on the number of consecutively
rejectable symbols.
Note that if the entire tape is rejected before the next ossifier arrives,
then there will be a collision between
a prepared leader and an ossifier,
as opposed to an ossifier hitting tape data
as in \figref{figSketchesZ}(z).

  The reader has probably noticed that we are encoding the tape as
  moving data.
  Although it might be more natural to encode it as tape data as described in \cite{Cook04a},
  encoding it as
  moving data gives us some nice computational simplifications:
  Only two slopes
  of glider appear in the initial \rst\ state, each glider
  only needs to be positioned relative to its immediate neighbors,
  and the moving data does not need to be reversed from the \cts\ tape.

The algorithm forced the \ts\ to have a size that is a multiple of six
so that every appendant in the \cts\ would have a length that is a multiple of six.
This is required so that the leader after a rejected appendant
will hit the next symbol of tape data correctly
(requiring the rejected appendant to have even length,
as shown in \figref{figSketchesVWXY}(y)),
producing invisibles that indeed pass through the ossifiers
(requiring the rejected appendant to have length a multiple of three,
as shown in \figref{figSketchesMNO}(o)).

A \cts\ appendant in this construction may have a length that is not a multiple of six
only if the \cts\ will always append that appendant.
Otherwise the construction will not work properly.
The safest approach is clearly to always use appendants whose length is a multiple of six.

%
%
%

\subsection{Converting back to a \tm}

We can continue with our series of conversions by converting the \rst\ initial condition
back into an initial state for a \tm\ tape.
The benefit obtained from this big cycle of conversions is that the final
\tm\ can be very small, since it only has to emulate \rst .
The program, on the other hand, will have become much larger
after going through the extensive compilation process in this cycle of conversions.
Here we will show some \tm s that can emulate \rst ,
and give an example for each of how it would emulate the evolution of the standard ether pattern.
Encoding the \rst\ inital state into an initial tape for these \tm s is fairly direct
and we will not discuss the details of these transformations.

   Figures~\ref{figTinyTM1} and~\ref{figTinyTM234}
   show the lookup tables for the \tm s of~\cite{Cook04a}.
   The captions give examples of initial tapes that
   allow the \tm s to implement \rst\ logic.
   Each machine emulates \rst\ by sweeping back and forth
   over an ever wider stretch of tape,
   with each rightward sweep computing one more
   step of \rst 's activity.

   More detailed initial tapes can be used to make the machines
   perform more detailed \rst\ computations, for example by
   inserting gliders into a central ether region,
   or including gliders in the periodic portion.
   Note that the periodic patterns on the sides of the \tm 's tape correspond to
   periodic paths through both space and time on the sides of the \rst\ evolution.

   In contrast with the \tm\ of \figref{figTinyTM1},
   the other three machines operate in such a way that the
   \rst\ rows shift one cell to the right on the \tm\ tape with each simulated time step.

More recently, Neary and Woods \cite{Neary07}
have achieved even smaller \tm s that similarly
emulate \rst .

\newcommand{\tmone}{
$
\begin{array}{r|c|c|c|c|c|l}
\multicolumn{1}{c}{}
& \multicolumn{1}{c}{0}
& \multicolumn{1}{c}{1}
& \multicolumn{1}{c}{0^2}
& \multicolumn{1}{c}{1^2}
& \multicolumn{1}{c}{\neq}
& \hspace{2 em}
\\ \cline{2-6}
S_0
& (0^2, S_0, L)
& (\neq, S_1, L)
& (0, S_0, R)
& (1, S_1, R)
& (1, S_0, R)
\\ \cline{2-6}
S_1
& (\neq, S_0, L)
& (1^2, S_1, L)
& X
& (0, S_1, R)
& (1, S_0, R)
\\ \cline{2-6}
\end{array}$
}

\newcommand{\tmtwo}{
$
\begin{array}{r|c|c|c|c|l}
\multicolumn{1}{c}{}
& \multicolumn{1}{c}{0_R}
& \multicolumn{1}{c}{1_R}
& \multicolumn{1}{c}{0_L}
& \multicolumn{1}{c}{1_L}
& \hspace{2 em}
\\ \cline{2-5}
S_{x0}
& (0_L, S_{x0}, R)
& (1_L, S_{01}, R)
& (0_R, S_{x0}, L)
& (1_R, S_{x0}, L)
\\ \cline{2-5}
S_{01}
& (1_L, S_{x0}, R)
& (1_L, S_{11}, R)
& X
& X
\\ \cline{2-5}
S_{11}
& (1_L, S_{x0}, R)
& (0_L, S_{11}, R)
& X
& X
\\ \cline{2-5}
\end{array}$
}

\newcommand{\tmthree}{
$
\begin{array}{r|c|c|c|l}
\multicolumn{1}{c}{}
& \multicolumn{1}{c}{0}
& \multicolumn{1}{c}{1}
& \multicolumn{1}{c}{B}
& \hspace{2 em}
\\ \cline{2-4}
S_{x0}
& (0, S_{x0}, R)
& (1, S_{01}, R)
& (0, S_B, L)
\\ \cline{2-4}
S_{01}
& (1, S_{x0}, R)
& (1, S_{11}, R)
& X
\\ \cline{2-4}
S_{11}
& (1, S_{x0}, R)
& (0, S_{11}, R)
& X
\\ \cline{2-4}
S_B
& (0, S_B, L)
& (1, S_B, L)
& (0, S_{x0}, R)
\\ \cline{2-4}
\end{array}$
}
\newcommand{\tmfour}{
$
\begin{array}{r|c|c|l}
\multicolumn{1}{c}{}
& \multicolumn{1}{c}{0}
& \multicolumn{1}{c}{1}
& \hspace{2 em}
\\ \cline{2-3}
S_{x0}
& (0, T_{x0}, R)
& (1, T_{01}, R)
\\ \cline{2-3}
S_{01}
& (1, T_{x0}, R)
& (1, T_{11}, R)
\\ \cline{2-3}
S_{11}
& (1, T_{x0}, R)
& (0, T_{11}, R)
\\ \cline{2-3}
S_{L}
& (0, T_{x0}, L)
& (1, T_{x0}, L)
\\ \cline{2-3}
T_{x0}
& (1, S_{x0}, R)
& (0, S_{L}, L)
\\ \cline{2-3}
T_{01}
& (1, S_{01}, R)
& X
\\ \cline{2-3}
T_{11}
& (1, S_{11}, R)
& X
\\ \cline{2-3}
\end{array}$
}

\begin{figure}
    \begin{center}
    \tmone
    \end{center}
 \smallcaption{\label{figTinyTM1}
   This 2 state 5 symbol \tm\ can be started in state $S_0$ on the two-way infinite tape
   $[0^{2}0\:1\:0]^{\infty}\fbox{$\!0^2\!$}\:[\neq\!1\!\neq\!1\:0^{2}\:0^{2}]^{\infty}$,
   starting on the cell marked here with a box,
   and it will start computing \rst 's ether pattern.
   This \tm\ is slightly modified from the one in~\cite{Cook04a}
   so that fewer transitions are needed (the unused one is marked with ``$X$'').
 }
\end{figure}

\begin{figure}
    \begin{center}
        \scalebox{.85}\tmtwo
    \end{center}
    \begin{center}
        \scalebox{.85}\tmthree
    \end{center}
    \begin{center}
        \scalebox{.85}\tmfour
    \end{center}
 \smallcaption{\label{figTinyTM234}
   The 3 state 4 symbol \tm\ can be started in state $S_{x0}$ on the 
   tape $[0_L 1_L 0_R]^\infty\fbox{$\!0_L\!$}[0_L 1_L 1_R 0_R 0_L]^\infty$
   to compute the ether pattern.
   The 4 state 3 symbol \tm\ can be started in state $S_{x0}$ on the
   tape $[B\:0\:1]^\infty\fbox{$\!\!B\!$}[B\:1\:1\:1\:1\:1\:0\:B]^\infty$
   to compute the ether pattern.
   The 7 state 2 symbol \tm\ can be started in state $S_{11}$ on the
   tape $[110011]^\infty\fbox{$\!0\!$}[1011010010]^\infty$,
   and it will compute the ether pattern on every second cell of the tape.
   This machine was compressed
   down to\label{EppsteinImprovement}
   seven states by David Eppstein~\cite{Eppstein98}.
 }
\end{figure}

\section{Towards a formal proof approach}

The original exposition of the \rst\ construction in~\cite{Cook04a}
took a motivated approach to understanding how
the construction could be made to work,
by showing how some parts of the construction placed
requirements on other parts of the construction,
and then showing how the other parts had enough
flexibility to satisfy those requirements.
Only the inner workings of the components and leaders
were treated as given without analysis.
Of course, those could have been explained as well,
as they were similarly designed by examining
how to satisfy the relevant constraints,
but this would have significantly lengthened
the exposition.

In the present paper we take the converse approach,
treating the entire construction as handed to us in a
completely specified form, as given in \secref{compiler},
for which all we have to do is check that
as \rst\ acts on the initial state,
all of the ensuing collisions will be of the right form.

Each collision,
whether between individual gliders or between clusters of gliders
(which can be thought of as very large gliders),
is completely determined by the spacing between
its parts.
In \rst 's two dimensional space-time,
spacings can be completely specified in terms of \ovd\ (``over distance'') and \upd\ (``up distance''),
described in~\cite{Cook04a}.

Our general approach is that
each time there is a collision, we can determine the spacing
between its parts by examining the spacings among
all of the previous collisions bordering the
ether region above the collision in question.
If every such region can be shown to lead to the proper collision,
then by induction, all the collisions will be correct.

All the collisions (at a glider cluster level) in this construction
are between \ebar\ material (leaders, table data, invisibles, moving data) on the right
and either $C_2$ material (tape data) or $A$ material (ossifiers, acceptors, rejectors) on the left.

The slopes of the $A$ gliders and $C$ gliders,
are directly related to the measurements that we need.
When $C_2$ material hits \ebar\ material,
the correctness of the collision depends only on the \ovd\ mod 4.
When $A$ material hits \ebar\ material,
the correctness of the collision depends only on the \upd\ mod 6.
The values are only important mod 4 and mod 6 because of the periodicity of the \ebar .

For most of the collisions,
some participants in the collision have emerged from crossing collisions.
In these cases, we measure the relevant distance back
at the point when those gliders were first created, before all the crossing collisions,
since the crossing collisions do not affect the relative distances.
Only collisions with acceptors and rejectors are never preceded by crossing collisions.

This construction uses only the following collisions:
\begin{itemize}
\item ossifiers hitting moving data or invisibles
  \begin{itemize}
  \item Measurements of these collisions are shown in \figref{figSketchesEFGH}.
  \item The moving data is created with the correct spacing as shown in \figref{figSketchesIJKL}.
  \item The invisibles are created with the correct spacing as shown in \figref{figSketchesMNO}.
  \end{itemize}
\item tape data passing through moving data or invisibles
  \begin{itemize}
  \item Measurements of these collisions are shown in \figref{figSketchesABCD}.
  \item The tape data is created with the correct spacing as shown in \figref{figSketchesEFGH}(e).
  \item The moving data and invisibles are created with the correct spacing as shown in \figref{figSketchesSTU}.
  \end{itemize}
\item an element of tape data hitting a prepared leader
  \begin{itemize}
  \item Measurements of this collision are shown in \figref{figSketchesVWXY}(v).
  \item The prepared leader is created with the correct spacing as shown in \figref{figSketchesVWXY}(w,x,y).
  \end{itemize}
\item an acceptor or rejector hitting table data
  \begin{itemize}
  \item Measurements for these collisions are shown in \figref{figSketchesIJKL}.
  \end{itemize}
\item an acceptor or rejector hitting a raw leader
  \begin{itemize}
  \item Measurements for these collisions are shown in \figref{figSketchesPQR}.
  \end{itemize}
\end{itemize}

We avoid an overly formal style in this presentation,
but the general approach used here
would be a good starting point
for a complete formalization of the proof, a formalization
of the sort that could be checked automatically by an automated proof checker.
Most of the simpler spacing claims have essentially already been checked insofar as
simulations of this construction appear to work with no problems.
But of course most of the more general claims that appear
cannot be verified in their full generality so trivially.

\ignore{
Since the first part of a short leader is exactly the same as the first
part of a regular leader, not only does its prepared form have
the same relationship to the absorbed rejector or acceptor
as a regular leader would (as shown in Figures~\ref{figSketchesMNO}(n,o)
and \ref{figSketchesPQR}(r)), but
the relationship between a prepared short
leader and its first invisible
is exactly the same as the relationship between a prepared regular leader
and its first invisible
(as shown in \figref{figSketchesMNO}(m) and used thereafter),
so the calculations showing that this invisible is spaced correctly
all apply to the short leader's first invisible as well.
}

\segsec{Variations on the Construction}

~

Other than the $A^4$\,\up{5} \ebar\ reaction used for ossification,
shown in \figref{figSketchesEFGH}(e),
there are two other reactions between an $A^4$ and an \ebar\ that result in a $C_2$.
One of them, in which the \ebar\ is \up{1} from the $A^4$, is clearly unusable
because it does not allow the ossification of \figref{figSketchesEFGH}(e) to work,
since the \ovd\ between the new $C_2$ and the \ebar\ is one, not zero.
The other, in which the \ebar\ is \up{2} from the $A^4$, does not have this problem.
If it were used, then the leaders and primary components
  would need to have different internal arrangements.
  It turns out that, regardless of the details of these arrangements,
  the equation of invisibility after rejection, $4+1+(2c-1)\cdot 5+2+5+5=0$,
  from \figref{figSketchesMNO}(m,o) and \figref{figSketchesEFGH}(g),
  which contains values that derive from nearly every aspect of the construction,
  would wind up being odd
  on one side and even on the other, and so the equation would be
  violated regardless of the value of $c$, and the construction
  would not work, regardless of any restrictions on the lengths of appendants.
  The construction as originally designed
  in February of 1994 suffered from this problem, which was discovered
  and fixed in the ensuing weeks.  Note that~\cite{Wolfram02} is
  mistaken when it says on page 1115
  that some mistakes in the proof were corrected in 1998.
  In fact, the basic construction, completed in 1994,
  leaves a lot of flexibility in choosing many of the
  spacings (even within the leaders and components, whose internal design is not
  discussed in~\cite{Cook04a}), and what happened in 1998 was that
  a particular set of arrangements were chosen (for compactness)
  while writing a program to automatically
  generate a \rst\ initial state
  corresponding to an arbitrary \cts\ (see \cite{Wolfram02}, page~1116).
  Finally, in 1999, a standardized set of methods of measurement
  and analysis were chosen during the
  writing of the exposition of the 1994 construction,
  the publication of which (in~\cite{Cook04a}) was delayed
  until after the publication of~\cite{Wolfram02}.

   The astute reader may notice that primary components
   (as well as short leaders)
   could be avoided if the acceptor and rejector could
   be produced \up{+2} from where the leader currently
   produces them.  The reader is invited to attempt this
   simplification, preferably without increasing
   the size of the construction.

\begin{figure}
 \centerline{
   \setlength{\unitlength}{1pt}\thicklines
   \beginpicture{330}{330}
    \put(0,210){\beginpicture{100}{100}\put(50,-15){\makebox(0,0){\sklab{a}}}
\put(90,90){\makebox(0,0){\ebar}}
\put(20,20){\line(1,1){60}}
\put(10,10){\makebox(0,0){\ebar}}
\put(45,10){\makebox(0,0){$C_2$}}
\put(45,20){\line(0,1){60}}
\put(45,90){\makebox(0,0){$C_2$}}
\multiput(66,66)(0,1.5){7}{\circle*{1}}
\put(60,77){\makebox(0,0){\ov{3}}}
\multiput(54,66)(0,1.5){7}{\circle*{1}}
\multiput(46.5,66)(2.5,0){3}{\overdots}
\multiput(46.5,25)(2.5,0){9}{\overdots}
\multiput(69,25)(0,1.5){11}{\circle*{1}}
\put(63,42){\makebox(0,0){\ov{0}}}
\multiput(57,57)(0,-1.5){14}{\circle*{1}}
\multiput(43.5,55)(-2.5,0){3}{\overdots}
\multiput(36,55)(0,1.5){7}{\circle*{1}}
\put(30,66){\makebox(0,0){\ov{3}}}
\multiput(24,24)(0,1.48){28}{\circle*{1}}
    \eendpicture}
    \put(160,210){\beginpicture{170}{120}\put(85,-15){\makebox(0,0){\sklab{b}}}
\put(80,10){\makebox(0,0){$C_2$}}
\put(80,20){\line(0,1){80}}
\put(80,110){\makebox(0,0){$C_2$}}
\put(110,110){\makebox(0,0){\ebar}}
\put(20,20){\line(1,1){80}}
\put(10,10){\makebox(0,0){\ebar}}
\put(160,110){\makebox(0,0){\ebar}}
\put(70,20){\line(1,1){80}}
\put(60,10){\makebox(0,0){\ebar}}
\multiput(99,60)(0,1.5){11}{\circle*{1}}
\multiput(81.5,60)(2.5,0){7}{\overdots}
\put(93,77){\makebox(0,0){\ov{0}}}
\multiput(87,87)(0,-1.5){11}{\circle*{1}}
\put(105,77){\makebox(0,0){\ov{3}}}
\multiput(111,61)(0,1.4){11}{\circle*{1}}
    \eendpicture}
    \put(0,20){\beginpicture{100}{100}\put(50,-15){\makebox(0,0){\sklab{c}}}
\put(90,90){\makebox(0,0){\ebar}}
\put(20,20){\line(1,1){60}}
\put(10,10){\makebox(0,0){\ebar}}
\put(38,10){\makebox(0,0){$C_2$}}
\put(38,20){\line(0,1){60}}
\put(38,90){\makebox(0,0){$C_2$}}
\put(62,10){\makebox(0,0){$C_2$}}
\put(62,20){\line(0,1){60}}
\put(62,90){\makebox(0,0){$C_2$}}
\multiput(50,50)(0,1.5){13}{\circle*{1}}
\put(44,70){\makebox(0,0){\ov{3}}}
\put(56,70){\makebox(0,0){\ov{3}}}
    \eendpicture}
    \put(140,20){\beginpicture{210}{150}\put(105,-15){\makebox(0,0){\sklab{d}}}
\put(70,140){\makebox(0,0){$C_2$}}
\put(70,130){\line(0,-1){70}}
\put(82,140){\makebox(0,0){$C_2$}}
\put(82,130){\line(0,-1){58}}
\put(94,140){\makebox(0,0){$C_2$}}
\put(94,130){\line(0,-1){46}}
\put(106,140){\makebox(0,0){$C_2$}}
\put(106,130){\line(0,-1){110}}
\put(106,10){\makebox(0,0){$C_2$}}
\put(140,140){\makebox(0,0){\ebar}}
\put(130,130){\line(-1,-1){110}}
\put(10,10){\makebox(0,0){\ebar}}
\put(160,140){\makebox(0,0){\ebar}}
\put(150,130){\line(-1,-1){50}}
\put(180,140){\makebox(0,0){\ebar}}
\put(170,130){\line(-1,-1){70}}
\put(200,140){\makebox(0,0){\ebar}}
\put(190,130){\line(-1,-1){90}}
\put(76,115){\makebox(0,0){\ov{2}}}
\put(88,115){\makebox(0,0){\ov{2}}}
\put(100,115){\makebox(0,0){\ov{2}}}
\put(112,127){\makebox(0,0){\ov{3}}}
\multiput(118,118)(0,1.5){6}{\circle*{1}}
\multiput(115,115)(0,-1.5){5}{\circle*{1}}
\put(121,114.3){\makebox(0,0){\ov{3}}}
\multiput(127,107)(0,1.5){5}{\circle*{1}}
\multiput(135,115)(0,-1.5){5}{\circle*{1}}
\put(141,114.3){\makebox(0,0){\ov{3}}}
\multiput(147,107)(0,1.5){5}{\circle*{1}}
\multiput(155,115)(0,-1.5){5}{\circle*{1}}
\put(161,114.3){\makebox(0,0){\ov{3}}}
\multiput(167,107)(0,1.5){5}{\circle*{1}}
\put(95,77){\makebox(0,0){?}}
\put(80,55){\makebox(0,0){?}}
\put(92,45){\makebox(0,0){?}}
    \eendpicture}
   \eendpicture
 }
 \smallcaption{\label{figSketchesABCD}
   An \ebar\ crossing a $C_2$.
   Vertical columns of dots represent the
   vertical columns of ether triangles used
   when measuring \ovd , which we treat as a value mod 4
   due to the periodicity of the \ebar .
   Sequences of parallel short vertical columns
   of dots represent the marking of every fourth
   column of ether triangles
   for convenience of measurement.
   {\bf (a)}%
~Some basic measurements of the unique collision in which an \ebar\ crosses a $C_2$.
   Note that the three measurements shown logically imply that the regenerated $C_2$
   must be \ovprot{2} from the regenerated \ebar .
   {\bf (b)}%
~Putting together some measurements from (a), we see that two consecutive
   \ebar s can both cross a $C_2$ if and only if they are \ovprot{3} from each other.
   Note that the spacing between the \ebar s will be the same after the collisions as
   before the collisions, since each \ebar\ undergoes an identical displacement in the collision.
   {\bf (c)}%
~Similarly, we see that consecutive $C_2$s can both cross an \ebar\ if and only if
   they are \ovprot{3+3}, which is \ovprot{2}, from each other.
   Note that they will still be \ovprot{2} from each other after the collisions,
   since they each undergo the same displacement.
   {\bf (d)}%
~If we align several $C_2$s and \ebar s so that the first collisions are
   crossing collisions, then what will the remaining collisions be?
   After crossing the first \ebar , the $C_2$s are still \ovprot{2} from each other,
   so the second \ebar\ will also cross them all, and indeed, all
   of the \ebar s will cross all of the $C_2$s.
 }
\end{figure}

\begin{figure}
 \centerline{
   \setlength{\unitlength}{1pt}\thicklines
   \beginpicture{360}{340}
    \put(0,200){\beginpicture{170}{140}\put(85,-15){\makebox(0,0){\sklab{e}}}
\put(10,130){\makebox(0,0){$A^4$}}
\put(20,120){\line(1,-1){45}}
\put(95,130){\makebox(0,0){$C_2$}}
\put(95,120){\line(0,-1){100}}
\put(95,10){\makebox(0,0){$C_2$}}
\put(120,130){\makebox(0,0){\ebar}}
\put(110,120){\line(-1,-1){45}}
\put(160,130){\makebox(0,0){\ebar}}
\put(150,120){\line(-1,-1){100}}
\put(40,10){\makebox(0,0){\ebar}}
\put(65,20){\line(0,1){55}}
\put(65,10){\makebox(0,0){$C_2$}}
\put(56,95){\makebox(0,0){\up{5}}}
\multiput(76,86)(-1,1){18}{\circle*{1}}
\multiput(104,114)(0,-1.5){10}{\circle*{1}}
\put(110,105.5){\makebox(0,0){\ov{3}}}
\multiput(116,86)(0,1.5){13}{\circle*{1}}
\put(110,95){\makebox(0,0){\ov{3}}}
\multiput(104,84)(0,1.5){7}{\circle*{1}}
\multiput(96.5,84)(2.5,0){3}{\overdots}
\multiput(74,84)(0,-1.5){10}{\circle*{1}}
\put(80,75.5){\makebox(0,0){\ov{3}}}
\multiput(86,56)(0,1.5){13}{\circle*{1}}
\put(80,65){\makebox(0,0){\ov{3}}}
\multiput(74,54)(0,1.5){7}{\circle*{1}}
\multiput(66.5,54)(2.5,0){3}{\overdots}
    \eendpicture}
    \put(210,200){\beginpicture{140}{100}\put(70,-15){\makebox(0,0){\sklab{f}}}
\put(8,90){\makebox(0,0){$A^4$}}
\put(18,80){\line(1,-1){60}}
\put(88,10){\makebox(0,0){$A^4$}}
\put(52,90){\makebox(0,0){$A^4$}}
\put(62,80){\line(1,-1){60}}
\put(132,10){\makebox(0,0){$A^4$}}
\put(110,90){\makebox(0,0){\ebar}}
\put(100,80){\line(-1,-1){60}}
\put(30,10){\makebox(0,0){\ebar}}
\put(50,59){\makebox(0,0){\up{0}}}
\multiput(70,50)(-1,1){18}{\circle*{1}}
\put(61,70){\makebox(0,0){\up{5}}}
    \eendpicture}
    \put(0,20){\beginpicture{140}{100}\put(70,-15){\makebox(0,0){\sklab{g}}}
\put(10,90){\makebox(0,0){$A^4$}}
\put(20,80){\line(1,-1){48}}
\put(90,90){\makebox(0,0){\ebar}}
\put(80,80){\line(-1,-1){60}}
\put(10,10){\makebox(0,0){\ebar}}
\put(130,90){\makebox(0,0){\ebar}}
\put(120,80){\line(-1,-1){48}}
\put(64,47){\makebox(0,0){\up{5}}}
\multiput(61,61)(1,-1){11}{\circle*{1}}
\multiput(69,69)(1,-1){13}{\circle*{1}}
\put(85,64){\makebox(0,0){\up{k}}}
\multiput(100,60)(-1,1){12}{\circle*{1}}
\put(70,26){\makebox(0,0){?}}
\put(90,30){\small $k=0$ for $C_2$}
\put(90,20){\small $k=1$ to cross}
    \eendpicture}
    \put(140,20){\beginpicture{220}{140}\put(120,-15){\makebox(0,0){\sklab{h}}}
\put(10,130){\makebox(0,0){$A^4$}}
\put(20,120){\line(1,-1){88}}
\put(100,130){\makebox(0,0){$A^4$}}
\put(110,120){\line(1,-1){25}}
\put(170,130){\makebox(0,0){\ebar}}
\put(160,120){\line(-1,-1){25}}
\put(210,130){\makebox(0,0){\ebar}}
\put(200,120){\line(-1,-1){88}}
\put(135,20){\line(0,1){75}}
\put(135,10){\makebox(0,0){$C_2$}}
\multiput(77.5,66.5)(2.5,2.5){5}{\updots}
\multiput(90,79)(-1,1){11}{\circle*{1}}
\put(88,92){\makebox(0,0){\up{5}}}
\multiput(101,90)(-1,1){11}{\circle*{1}}
\multiput(103.5,92.5)(2.5,2.5){7}{\updots}
\put(127,99){\updots}
\multiput(122.5,98.5)(1,-1){10}{\circle*{1}}
\put(135,90){\squiggle}
\multiput(146.5,83)(-1,1){9}{\circle*{1}}
\put(150,90){\makebox(0,0){\up{4}}} 
\multiput(157.5,93.5)(-1,1){12}{\circle*{1}}
\put(161,101){\makebox(0,0){\up{k}}}
\multiput(164.5,108.5)(1,-1){12}{\circle*{1}}
\multiput(146.5,68)(-1,1){9}{\circle*{1}}
\put(135,75){\squiggle}
\multiput(131.5,74)(-1,1){10}{\circle*{1}}
\multiput(122,78.5)(-2.5,-2.5){8}{\updots}
\multiput(100,60.5)(1,-1){13}{\circle*{1}}
\put(116,56){\makebox(0,0){\up{4}}}
\multiput(119.5,63.5)(1,-1){12}{\circle*{1}}
\multiput(157,114)(2.5,0){8}{\overdots}
\multiput(177,114)(0,1.5){7}{\circle*{1}}
\put(183,125){\makebox(0,0){\ov{3}}}
\multiput(189,109)(0,1.55){10}{\circle*{1}}
\put(110,26){\makebox(0,0){?}}
\put(150,40){\small $k=4$ for $C_2$}
\put(150,30){\small $k=5$ to cross}
    \eendpicture}
   \eendpicture
 }
 \smallcaption{\label{figSketchesEFGH}
   How an \ebar\ can hit an $A^4$.
   Vertical columns of dots are as in \figref{figSketchesABCD}.
   Diagonal rows of dots represent the diagonal
   rows of ether triangles used when measuring \upd ,
   which we treat as a value mod 6 due to the periodicity of the \ebar .
   Sequences of parallel short diagonal rows of dots
   represent the marking of every sixth row of ether triangles
   for convenience of measurement.
   {\bf (e)}
~There are three ways an \ebar\ can hit an $A^4$ to produce a $C_2$,
   but only one works for the construction.  It is the one
   where the \ebar\ is \up{5} from the $A^4$.
   The resulting $C_2$ is \ovprot{0} from the \ebar\ that created it.
   {\bf (f)}
~An \ebar\ will cross an $A^4$ if and only if
   it is \up{0} from the $A^4$.
   The crossing causes a visually striking but irrelevant
   displacement (not indicated here) to the $A^4$.
   The incoming $A^4$ is \up{5} from the regenerated \ebar .
   We see that consecutive $A^4$s must be \up{5} from each
   other if they are both to cross an \ebar ,
   and the construction always uses such a spacing between $A^4$s.
   Every \ebar\ will either pass through all the $A^4$s
   (we call such an \ebar\ an ``invisible''),
   or else it is ``moving data'' and will be converted (``ossified'')
   into a $C_2$ (``tape data'') by the first $A^4$ (``ossifier'') it hits, as in (e).
   {\bf (g)}
~After an invisible \ebar , the \upd\ to the next \ebar\ determines
   whether it is another invisible or ossifiable moving data.
   {\bf (h)}
~After a moving data \ebar , the \upd\ to the next \ebar\ determines
   whether it is more moving data or an invisible.
 }
\end{figure}

\begin{figure}
 \centerline{
   \setlength{\unitlength}{1pt}\thicklines
   \beginpicture{320}{260}
    \put(0,20){\beginpicture{140}{100}\put(70,-15){\makebox(0,0){\sklab{k}}}
\put(30,90){\grAcc}
\put(40,80){\line(1,-1){17}}
\figdot{60}{60}{figComponents}
\put(63,57){\line(1,-1){14}}
\figdot{80}{40}{figComponents}
\put(83,37){\line(1,-1){17}}
\put(110,10){\grAcc}
\put(10,10){\grMd}
\put(20,20){\line(1,1){37}}
\put(80,80){\line(-1,-1){17}}
\put(90,90){\grPriStd}
\put(50,10){\grMd}
\put(60,20){\line(1,1){17}}
\put(120,80){\line(-1,-1){37}}
\put(130,90){\grStd}
\multiput(75,49)(2.5,2.5){6}{\updots}
\multiput(90,64)(-1,1){11}{\circle*{1}}
\put(76,67){\makebox(0,0){\up{3}}}
\multiput(66,66)(1,-1){7}{\circle*{1}}
\multiput(90,64)(1,-1){4}{\circle*{1}}
\put(97,68){\makebox(0,0){\up{5}}}
\multiput(101,75)(1,-1){8}{\circle*{1}}
\multiput(50,36)(-1,1){5}{\circle*{1}}
\put(42,33){\makebox(0,0){\up{0}}}
\multiput(38,26)(-1,1){7}{\circle*{1}}
\multiput(67,49)(-2.5,-2.5){6}{\updots}
\multiput(50,36)(1,-1){10}{\circle*{1}}
\put(63,34){\makebox(0,0){\up{4}}}
\multiput(74,34)(-1,1){8}{\circle*{1}}
    \eendpicture}
    \put(180,20){\beginpicture{140}{100}\put(70,-15){\makebox(0,0){\sklab{l}}}
\put(30,90){\grRej}
\put(40,80){\line(1,-1){17}}
\figdot{60}{60}{figComponents}
\put(63,57){\line(1,-1){14}}
\figdot{80}{40}{figComponents}
\put(83,37){\line(1,-1){17}}
\put(110,10){\grRej}
\put(63,63){\line(1,1){17}}
\put(90,90){\grPriStd}
\put(83,43){\line(1,1){37}}
\put(130,90){\grStd}
\multiput(75,49)(2.5,2.5){6}{\updots}
\multiput(90,64)(-1,1){11}{\circle*{1}}
\put(76,67){\makebox(0,0){\up{0}}}
\multiput(66,66)(1,-1){7}{\circle*{1}}
\multiput(90,64)(1,-1){4}{\circle*{1}}
\put(97,68){\makebox(0,0){\up{2}}}
\multiput(101,75)(1,-1){8}{\circle*{1}}
    \eendpicture}
    \put(0,160){\beginpicture{140}{100}\put(70,-15){\makebox(0,0){\sklab{i}}}
\put(60,90){\grTd}
\put(60,80){\line(0,-1){16}}
\figdot{60}{60}{figLeaders}
\put(63,57){\line(1,-1){14}}
\figdot{80}{40}{figComponents}
\put(83,37){\line(1,-1){17}}
\put(110,10){\grAcc}
\put(10,10){\grInv}
\put(20,20){\line(1,1){37}}
\put(80,80){\line(-1,-1){17}}
\put(90,90){\grPrep}
\put(50,10){\grMd}
\put(60,20){\line(1,1){17}}
\put(120,80){\line(-1,-1){37}}
\put(130,90){\grPri}
\multiput(75,49)(2.5,2.5){6}{\updots}
\multiput(90,64)(-1,1){11}{\circle*{1}}
\put(76,67){\makebox(0,0){\up{3}}}
\multiput(66,66)(1,-1){7}{\circle*{1}}
\multiput(90,64)(1,-1){4}{\circle*{1}}
\put(97,68){\makebox(0,0){\up{5}}}
\multiput(101,75)(1,-1){8}{\circle*{1}}
\multiput(50,36)(-1,1){5}{\circle*{1}}
\put(42,33){\makebox(0,0){\up{0}}}
\multiput(38,26)(-1,1){7}{\circle*{1}}
\multiput(67,49)(-2.5,-2.5){6}{\updots}
\multiput(50,36)(1,-1){10}{\circle*{1}}
\put(63,34){\makebox(0,0){\up{0}}}
\multiput(74,34)(-1,1){8}{\circle*{1}}
    \eendpicture}
    \put(180,160){\beginpicture{140}{100}\put(70,-15){\makebox(0,0){\sklab{j}}}
\put(60,90){\grTd}
\put(60,80){\line(0,-1){16}}
\figdot{60}{60}{figLeaders}
\put(63,57){\line(1,-1){14}}
\figdot{80}{40}{figComponents}
\put(83,37){\line(1,-1){17}}
\put(110,10){\grRej}
\put(10,10){\grInv}
\put(20,20){\line(1,1){37}}
\put(80,80){\line(-1,-1){17}}
\put(90,90){\grPrep}
\put(120,80){\line(-1,-1){37}}
\put(130,90){\grPri}
\multiput(75,49)(2.5,2.5){6}{\updots}
\multiput(90,64)(-1,1){11}{\circle*{1}}
\put(76,67){\makebox(0,0){\up{0}}}
\multiput(66,66)(1,-1){7}{\circle*{1}}
\multiput(90,64)(1,-1){4}{\circle*{1}}
\put(97,68){\makebox(0,0){\up{2}}}
\multiput(101,75)(1,-1){8}{\circle*{1}}
    \eendpicture}
   \eendpicture
 }
 \smallcaption{\label{figSketchesIJKL}
   Primary ({\grs{pri}{PRI}})
   and standard ({\grs{std}{STD}})
   components will get processed correctly by
   either an acceptor ({\grs{acc}{ACC}}) or a rejector ({\grs{rej}{REJ}}).
   An acceptor converts components into moving data ({\grs{md}{MD}}),
   whereas a rejector deletes components.
   The acceptor or rejector is produced by a prepared
   leader ({\grs{prep}{PREP}}) hitting a character of tape data ({\grs{td}{TD}}),
   which produces a pair of invisibles ({\grs{inv}{INV}}) as well as
   the acceptor or rejector.
   Note that lines in these diagrams represent clusters
   of parallel gliders, and the collisions between clusters
   are marked with a 
   circle representing the many
   collisions that occur where the clusters meet.
   A measurement to or from a cluster is made to or from
   the closest glider in the cluster, namely the one that touches
   the ether in which the measurement is being made.
 }
\end{figure}

\begin{figure}
 \centerline{
   \setlength{\unitlength}{1pt}\thicklines
   \beginpicture{360}{350}
    \put(0,230){\beginpicture{160}{120}\put(70,-15){\makebox(0,0){\sklab{m}}}
\put(80,110){\grTd}
\put(80,100){\line(0,-1){16}}
\figdot{80}{80}{figCsEs}
\put(80,76){\line(0,-1){32}}
\figdot{80}{40}{figLeaders}
\put(83,37){\line(1,-1){17}}
\put(110,10){\grAccRej}
\put(10,10){\grInvMd}
\put(20,20){\line(1,1){57}}
\put(100,100){\line(-1,-1){17}}
\put(110,110){\grInvMd}
\put(50,10){\grInv}
\put(60,20){\line(1,1){17}}
\put(140,100){\line(-1,-1){57}}
\put(150,110){\grPrep}
\multiput(72.5,72.5)(1,-1){5}{\circle*{1}}
\put(80,69.5){\squiggle}
\multiput(83.5,70.5)(1,-1){7}{\circle*{1}}
\put(93,72){\makebox(0,0){\up{2}}}
\multiput(100,76)(-1,1){13}{\circle*{1}}
\put(104,83){\makebox(0,0){\up{k}}}
\multiput(108,90)(1,-1){12}{\circle*{1}}
\multiput(87.5,47.5)(-1,1){5}{\circle*{1}}
\put(80,50.5){\squiggle}
\multiput(76.5,49.5)(-1,1){7}{\circle*{1}}
\multiput(70,51)(-2.5,-2.5){7}{\updots}
\multiput(50.5,35.5)(1,-1){9}{\circle*{1}}
\put(63,34){\makebox(0,0){\up{3}}}
\multiput(67,41)(1,-1){8}{\circle*{1}}
    \eendpicture}
    \put(200,230){\beginpicture{160}{120}\put(70,-15){\makebox(0,0){\sklab{n}}}
\put(50,110){\grAcc}
\put(60,100){\line(1,-1){17}}
\figdot{80}{80}{figComponents}
\put(83,77){\line(1,-1){14}}
\figdot{100}{60}{figAbsorption}
\put(10,10){\grMd}
\put(20,20){\line(1,1){57}}
\put(100,100){\line(-1,-1){17}}
\put(110,110){\grStd}
\put(50,10){\grPrep}
\put(60,20){\line(1,1){37}}
\put(140,100){\line(-1,-1){37}}
\put(150,110){\grRaw}
\multiput(70,56)(-1,1){5}{\circle*{1}}
\put(62,53){\makebox(0,0){\up{0}}}
\multiput(58,46)(-1,1){7}{\circle*{1}}
\multiput(87,69)(-2.5,-2.5){6}{\updots}
\multiput(70,56)(1,-1){10}{\circle*{1}}
\put(83,54){\makebox(0,0){\up{0}}}
\multiput(94,54)(-1,1){8}{\circle*{1}}
\multiput(43,43)(1,-1){16}{\circle*{1}}
\put(62,35){\makebox(0,0){\up{k}}}
\multiput(66,42)(1,-1){9}{\circle*{1}}
\put(90,27){\small $k=0$}
    \eendpicture}
    \put(25,20){\beginpicture{310}{170}\put(155,-15){\makebox(0,0){\sklab{o}}}
\put(130,160){\grTd}
\put(130,150){\line(0,-1){16}}
\figdot{130}{130}{figLeaders}
\multiput(133,127)(.1,-.1){41}{\circle*{.7}} 
\figdot{140}{120}{figComponents}
\put(143,117){\line(1,-1){19}}
\figdot{165}{95}{figComponents}
\put(168,92){\line(1,-1){29}}
\figdot{200}{60}{figAbsorption}
\put(190,85){\grRej}
\put(10,10){\grInv}
\put(20,20){\line(1,1){107}}
\put(133,133){\line(1,1){17}}
\put(157,160){\grPrep} 
\put(180,160){\grPri}
\put(143,123){\line(1,1){27}}
\put(230,160){\grStd}
\put(168,98){\line(1,1){52}}
\put(300,160){\grRaw}
\put(160,20){\line(1,1){37}}
\put(203,63){\line(1,1){87}}
\put(150,10){\grPrep}
\multiput(115,101)(-1,1){8}{\circle*{1}}
\put(119,108){\makebox(0,0){\up{4}}}
\multiput(135,123)(-2.5,-2.5){4}{\updots} 
\multiput(123,115)(1,-1){13}{\circle*{1}}
\put(139,110){\makebox(0,0){\up{1}}}
\multiput(160,98)(-2.5,-2.5){4}{\updots} 
\multiput(148,90)(1,-1){13}{\circle*{1}}
\put(164,85){\makebox(0,0){\up{5}}}
{\thinlines
\put(149,109){\line(-1,-3){14}}
\put(173,85){\line(-1,-3){6}}
}
\put(150.8,60){\makebox(0,0){$\stackrel{\underbrace{\hspace{31.5pt}}}{2c-1}$}}
\multiput(187,69)(-2.5,-2.5){6}{\updots}
\multiput(170,56)(1,-1){10}{\circle*{1}}
\put(183,54){\makebox(0,0){\up{2}}}
\multiput(194,54)(-1,1){8}{\circle*{1}}
\multiput(88,88)(1,-1){61}{\circle*{1}}
\put(152,35){\makebox(0,0){\up{k}}}
\multiput(156,42)(1,-1){14}{\circle*{1}}
\put(190,22){\small $k=2$}
    \eendpicture}
   \eendpicture
 }
 \smallcaption{\label{figSketchesMNO}
   Invisibles are the result of a prepared leader hitting an element of tape data.
   {\bf (m)}%
~The alignment of the invisibles ($k+5$) depends on the alignment ($k$) of the prepared leader.
   The incoming invisible or moving data was originally produced as in (o) or (n).
   {\bf (n)}%
~When invisibles follow moving data,
   the leader was prepared by an acceptor,
   and the prepared leader's offset from the moving data is $k=0$.
   As shown in (m),
   this gives the invisibles an alignment of $k+5=5$,
   which will yield the correct interaction with the ossifiers,
   as shown in \figref{figSketchesEFGH}(h).
   {\bf (o)}%
~When invisibles follow previous invisibles,
   the leader was prepared by a rejector.
   If $c$ characters of table data were rejected,
   then $k=4+1+(2c-1)\cdot 5+2=10c+2$.
   Since $c$ is a multiple of 6, we get $k=2$.
   In the special case $c=0$,
   the previous leader was a short leader,
   and this figure does not apply, but we still get
   $k=2$ as shown in \figref{figSketchesPQR}(r).
   Either way, the invisibles are produced as in (m) with an alignment of $k+5=1$,
   yielding the correct collision with the ossifiers
   as shown in \figref{figSketchesEFGH}(g).
 }
\end{figure}

\begin{figure}
 \centerline{
   \setlength{\unitlength}{1pt}\thicklines
   \beginpicture{320}{280}
    \put(0,180){\beginpicture{140}{100}\put(70,-15){\makebox(0,0){\sklab{p}}}
\put(30,90){\grAcc}
\put(40,80){\line(1,-1){17}}
\figdot{60}{60}{figComponents}
\put(63,57){\line(1,-1){14}}
\figdot{80}{40}{figAbsorption}
\put(10,10){\grMd}
\put(20,20){\line(1,1){37}}
\put(80,80){\line(-1,-1){17}}
\put(90,90){\grStd}
\put(50,10){\grPrep}
\put(60,20){\line(1,1){17}}
\put(120,80){\line(-1,-1){37}}
\put(130,90){\grRaw}
\multiput(75,49)(2.5,2.5){6}{\updots}
\multiput(90,64)(-1,1){11}{\circle*{1}}
\put(76,67){\makebox(0,0){\up{3}}}
\multiput(66,66)(1,-1){7}{\circle*{1}}
\multiput(90,64)(1,-1){4}{\circle*{1}}
\put(97,68){\makebox(0,0){\up{3}}}
\multiput(101,75)(1,-1){8}{\circle*{1}}
    \end{picture}}
    \put(180,180){\beginpicture{140}{100}\put(70,-15){\makebox(0,0){\sklab{q}}}
\put(30,90){\grRej}
\put(40,80){\line(1,-1){17}}
\figdot{60}{60}{figComponents}
\put(63,57){\line(1,-1){14}}
\figdot{80}{40}{figAbsorption}
\put(80,80){\line(-1,-1){17}}
\put(90,90){\grStd}
\put(50,10){\grPrep}
\put(60,20){\line(1,1){17}}
\put(120,80){\line(-1,-1){37}}
\put(130,90){\grRaw}
\multiput(75,49)(2.5,2.5){6}{\updots}
\multiput(90,64)(-1,1){11}{\circle*{1}}
\put(76,67){\makebox(0,0){\up{0}}}
\multiput(66,66)(1,-1){7}{\circle*{1}}
\multiput(90,64)(1,-1){4}{\circle*{1}}
\put(97,68){\makebox(0,0){\up{0}}}
\multiput(101,75)(1,-1){8}{\circle*{1}}
    \eendpicture}
    \put(80,20){\beginpicture{160}{120}\put(80,-15){\makebox(0,0){\sklab{r}}}
\put(80,110){\grTd}
\put(80,100){\line(0,-1){16}}
\figdot{80}{80}{figShortLeader}
\put(83,77){\line(1,-1){14}}
\figdot{100}{60}{figAbsorption}
\put(10,10){\grInv}
\put(20,20){\line(1,1){57}}
\put(100,100){\line(-1,-1){17}}
\put(110,110){\grShortPrep}
\put(50,10){\grPrep}
\put(60,20){\line(1,1){37}}
\put(140,100){\line(-1,-1){37}}
\put(150,110){\grRaw}
\multiput(70,56)(-1,1){5}{\circle*{1}}
\put(62,53){\makebox(0,0){\up{0}}}
\multiput(58,46)(-1,1){7}{\circle*{1}}
\multiput(87,69)(-2.5,-2.5){6}{\updots}
\multiput(70,56)(1,-1){10}{\circle*{1}}
\put(83,54){\makebox(0,0){\up{2}}}
\multiput(94,54)(-1,1){8}{\circle*{1}}
\multiput(43,43)(1,-1){16}{\circle*{1}}
\put(62,35){\makebox(0,0){\up{k}}}
\multiput(66,42)(1,-1){9}{\circle*{1}}
\multiput(95,69)(2.5,2.5){6}{\updots}
\multiput(110,84)(-1,1){11}{\circle*{1}}
\put(96,87){\makebox(0,0){\up{0}}}
\multiput(86,86)(1,-1){7}{\circle*{1}}
\multiput(110,84)(1,-1){4}{\circle*{1}}
\put(117,88){\makebox(0,0){\up{0}}}
\multiput(121,95)(1,-1){8}{\circle*{1}}
\put(85,25){\small $k=2$}
    \eendpicture}
   \eendpicture
 }
 \smallcaption{\label{figSketchesPQR}
   Raw leaders are placed \up{0} from the previous component,
   which yields the correct alignment when prepared by
   either an acceptor (p) or a rejector (q).
   If there were no previous components, then the raw leader
   is placed \up{0} from the previous short leader (r)
   (whose lower right gliders are untouched by the preparation),
   again yielding the correct alignment for preparation.
   Raw leaders that come after short leaders
   yield prepared leaders aligned as in (r), with $k=2$.
   This is the zero-component version of \figref{figSketchesMNO}(o).
   Note that raw {\em short}\/ leaders,
   due to the different position of their initial \ebar , are always placed \up{+3}
   higher, as measured through the $E^n$s,
   than the raw regular leaders shown here.
 }
\end{figure}

\begin{figure}
 \centerline{
   \setlength{\unitlength}{1pt}\thicklines
   \beginpicture{360}{320}
    \put(-10,190){\beginpicture{220}{130}\put(110,-15){\makebox(0,0){\sklab{s}}}
\put(90,120){\grTd}
\put(90,110){\line(0,-1){16}}
\figdot{90}{90}{figLeaders}
\put(93,87){\line(1,-1){39}}
\figdot{135}{45}{figComponents}
\put(138,42){\line(1,-1){23}}
\put(170,10){\grAcc}
\put(10,10){\grInv}
\put(20,20){\line(1,1){67}}
\put(110,110){\line(-1,-1){17}}
\put(120,120){\grPrep}
\put(100,10){\grMd}
\put(110,20){\line(1,1){22}}
\put(200,110){\line(-1,-1){62}}
\put(210,120){\grPri}
\multiput(86,86)(0,-1.5){13}{\circle*{1}}
\put(92,69.5){\makebox(0,7){\ov{1}}}
\multiput(98,68)(0,1.5){8}{\circle*{1}}
\put(97.5,82.5){\vsquiggle}
\multiput(97,86.5)(0,1.5){8}{\circle*{1}}
\multiput(149,91)(0,-1.5){22}{\circle*{1}}
\multiput(111.5,82)(2.5,0){15}{\overdots}
\put(103,89){\makebox(0,7){\ov{0}}}
\multiput(109,91)(0,-1.5){12}{\circle*{1}}
\put(109.5,70.5){\vsquiggle}
\multiput(110,66.5)(0,-1.5){20}{\circle*{1}}
\put(116,39.5){\makebox(0,7){\ov{2}}}
\multiput(122,41)(0,-1.5){7}{\circle*{1}}
    \eendpicture}
    \put(70,20){\beginpicture{220}{130}\put(110,-15){\makebox(0,0){\sklab{u}}}
\put(60,120){\grAcc}
\put(70,110){\line(1,-1){17}}
\figdot{90}{90}{figComponents}
\put(93,87){\line(1,-1){39}}
\figdot{135}{45}{figComponents}
\put(138,42){\line(1,-1){23}}
\put(170,10){\grAcc}
\put(10,10){\grMd}
\put(20,20){\line(1,1){67}}
\put(110,110){\line(-1,-1){17}}
\put(120,120){\grPriStd}
\put(100,10){\grMd}
\put(110,20){\line(1,1){22}}
\put(200,110){\line(-1,-1){62}}
\put(210,120){\grStd}
\multiput(86,86)(0,-1.5){13}{\circle*{1}}
\put(92,69.5){\makebox(0,7){\ov{1}}}
\multiput(98,68)(0,1.5){8}{\circle*{1}}
\put(97.5,82.5){\vsquiggle}
\multiput(97,86.5)(0,1.5){8}{\circle*{1}}
\multiput(149,91)(0,-1.5){22}{\circle*{1}}
\multiput(111.5,82)(2.5,0){15}{\overdots}
\put(103,89){\makebox(0,7){\ov{0}}}
\multiput(109,91)(0,-1.5){12}{\circle*{1}}
\put(109.5,70.5){\vsquiggle}
\multiput(110,66.5)(0,-1.5){20}{\circle*{1}}
\put(116,39.5){\makebox(0,7){\ov{2}}}
\multiput(122,41)(0,-1.5){7}{\circle*{1}}
    \eendpicture}
    \put(200,190){\beginpicture{160}{120}\put(80,-15){\makebox(0,0){\sklab{t}}}
\put(39,110){\grTd}
\put(39,100){\line(0,-1){57}}
\put(39,20){\line(0,1){15}}
\put(39,10){\grTd}
\put(80,110){\grTd}
\put(80,100){\line(0,-1){16}}
\put(83,77){\line(1,-1){57}}
\put(150,10){\grAccRej}
\put(110,110){\grPrep}
\put(100,100){\line(-1,-1){17}}
\figdot{80}{80}{figLeaders}
\put(77,77){\line(-1,-1){35}}
\put(75,62){\grInv}
\figdot{39}{39}{figCsEs}
\put(20,20){\line(1,1){16}}
\put(10,10){\grInv}
\put(45,75){\makebox(0,7){\ov{2}}}
\multiput(51,73.5)(0,1.5){14}{\circle*{1}}
\multiput(53.5,90)(2.5,0){11}{\overdots}
\put(57,75){\makebox(0,7){\ov{1}}}
\multiput(63,76.5)(0,-1.5){10}{\circle*{1}}
    \eendpicture}
   \eendpicture
 }
 \smallcaption{\label{figSketchesSTU}
   Moving data and invisibles are created with the correct alignment
   to pass through tape data.
   The top collision in (s) is also the top collision in (t),
   but (t) applies even when a rejector is produced.
   The lower collision in (s) is also the top collision in (u),
   but (u) applies also to later components, where the lower collision in (u)
   becomes the top collision in (u).
   ~{\bf (s)}
~The first moving data after an invisible has correct alignment to pass
   through tape data, as in \figref{figSketchesABCD}(b).
   ~{\bf (t)}
~An invisible always has correct alignment to pass through tape data,
   as in \figref{figSketchesABCD}(a,c).
   ~{\bf (u)}
~Moving data after previous moving data is also aligned correctly
   for passing through tape data, as in \figref{figSketchesABCD}(b).
 }
\end{figure}

\begin{figure}
 \centerline{
   \setlength{\unitlength}{1pt}\thicklines
   \beginpicture{360}{500}
    \put(90,370){\beginpicture{180}{130}\put(90,-15){\makebox(0,0){\sklab{v}}}
\put(90,120){\grTd}
\put(90,110){\line(0,-1){16}}
\figdot{90}{90}{figCsEs}
\put(90,86){\line(0,-1){42}}
\figdot{90}{40}{figLeaders}
\put(93,37){\line(1,-1){17}}
\put(120,10){\grAccRej}
\put(120,120){\makebox(0,0){\ebar}}
\put(110,110){\line(-1,-1){17}}
\put(87,87){\line(-1,-1){67}}
\put(10,10){\makebox(0,0){\ebar}}
\put(170,120){\grPrep}
\put(160,110){\line(-1,-1){67}}
\put(87,37){\line(-1,-1){17}}
\put(60,10){\grInv}
\multiput(97,97)(0,-1.5){12}{\circle*{1}}
\put(103,82){\makebox(0,7){\ov{0}}}
\multiput(109,83.5)(0,-1.5){11}{\circle*{1}}
\multiput(106.5,68.5)(-2.5,0){7}{\overdots}
\put(115,82){\makebox(0,7){\ov{1}}}
\multiput(121,83.5)(0,-1.5){9}{\circle*{1}}
    \eendpicture}
    \put(0,220){\beginpicture{170}{110}\put(80,-15){\makebox(0,0){\sklab{w}}}
\put(70,100){\grTd}
\put(70,90){\line(0,-1){16}}
\figdot{70}{70}{figShortLeader}
\put(73,67){\line(1,-1){24}}
\figdot{100}{40}{figAbsorption}
\put(100,100){\grShortPrep}
\put(90,90){\line(-1,-1){17}}
\put(67,67){\line(-1,-1){47}}
\put(10,10){\grInv}
\put(160,100){\grRaw}
\put(150,90){\line(-1,-1){47}}
\put(97,37){\line(-1,-1){17}}
\put(70,10){\grPrep}
\multiput(66,66)(0,-1.5){13}{\circle*{1}}
\put(72,49.5){\makebox(0,7){\ov{2}}}
\multiput(78,37.5)(0,1.5){15}{\circle*{1}}
\put(77.5,62.5){\vsquiggle}
\multiput(77,66.5)(0,1.5){8}{\circle*{1}}
\put(84,39){\makebox(0,7){\ov{3}}}
\multiput(90,40.5)(0,-1.5){8}{\circle*{1}}
    \eendpicture}
    \put(190,220){\beginpicture{170}{110}\put(80,-15){\makebox(0,0){\sklab{x}}}
\put(40,100){\grAcc}
\put(50,90){\line(1,-1){17}}
\figdot{70}{70}{figComponents}
\put(73,67){\line(1,-1){24}}
\figdot{100}{40}{figAbsorption}
\put(100,100){\grStd}
\put(90,90){\line(-1,-1){17}}
\put(67,67){\line(-1,-1){47}}
\put(10,10){\grMd}
\put(160,100){\grRaw}
\put(150,90){\line(-1,-1){47}}
\put(97,37){\line(-1,-1){17}}
\put(70,10){\grPrep}
\multiput(66,66)(0,-1.5){13}{\circle*{1}}
\put(72,49.5){\makebox(0,7){\ov{1}}}
\multiput(78,37.5)(0,1.5){15}{\circle*{1}}
\put(77.5,62.5){\vsquiggle}
\multiput(77,66.5)(0,1.5){8}{\circle*{1}}
\put(84,39){\makebox(0,7){\ov{0}}}
\multiput(90,40.5)(0,-1.5){8}{\circle*{1}}
    \eendpicture}
    \put(20,20){\beginpicture{320}{160}\put(160,-15){\makebox(0,0){\sklab{y}}}
\put(120,150){\grTd}
\put(120,140){\line(0,-1){16}}
\figdot{120}{120}{figLeaders}
\put(123,117){\line(1,-1){18}}
\figdot{144}{96}{figComponents}
\put(147,93){\line(1,-1){18}}
\figdot{168}{72}{figComponents}
\put(171,69){\line(1,-1){26}}
\figdot{200}{40}{figAbsorption}
\put(190,65){\grRej}
\put(10,10){\grInv}
\put(20,20){\line(1,1){97}}
\put(123,123){\line(1,1){17}}
\put(150,150){\grPrep}
\put(198,150){\grPri}
\put(147,99){\line(1,1){41}}
\put(246,150){\grStd}
\put(171,75){\line(1,1){65}}
\put(170,10){\grPrep}
\put(180,20){\line(1,1){17}}
\put(203,43){\line(1,1){97}}
\put(310,150){\grRaw}
\multiput(114.5,114.5)(0,-1.5){12}{\circle*{1}}
\put(120.5,99.5){\makebox(0,7){\ov{1}}}
\multiput(126.5,110)(0,-1.5){17}{\circle*{1}}
\put(126,114){\vsquiggle}
\multiput(125.5,118)(0,1.5){6}{\circle*{1}}
\put(132.5,87.5){\makebox(0,7){\ov{0}}}
\multiput(138.5,98)(0,-1.5){17}{\circle*{1}}
\put(138,102){\vsquiggle}
\multiput(137.5,106)(0,1.5){9}{\circle*{1}}
\multiput(137.5,118)(2.5,0){8}{\overdots}
\multiput(155,121)(0,-1.5){10}{\circle*{1}}
\put(144.5,75.5){\makebox(0,7){\ov{3}}}
\multiput(150.5,86)(0,-1.5){17}{\circle*{1}}
\put(150,90){\vsquiggle}
\multiput(149.5,94)(0,1.5){6}{\circle*{1}}
\put(156.5,63.5){\makebox(0,7){\ov{0}}}
\multiput(162.5,74)(0,-1.5){17}{\circle*{1}}
\put(162,78){\vsquiggle}
\multiput(161.5,82)(0,1.5){9}{\circle*{1}}
\multiput(161.5,94)(2.5,0){8}{\overdots}
\multiput(179,97)(0,-1.5){10}{\circle*{1}}
\put(168.5,51.5){\makebox(0,7){\ov{3}}}
\multiput(174.5,62)(0,-1.5){17}{\circle*{1}}
\put(174,66){\vsquiggle}
\multiput(173.5,70)(0,1.5){6}{\circle*{1}}
\put(180.5,39.5){\makebox(0,7){\ov{0}}}
\multiput(186.5,41)(0,-1.5){10}{\circle*{1}}
{\thinlines
\put(150,69){\line(-1,-3){12}}
\put(174,45){\line(-1,-3){4}}
}
\put(153.8,26){\makebox(0,0){$\stackrel{\underbrace{\hspace{31.5pt}}}{2c-1}$}}
    \eendpicture}
   \eendpicture
 }
 \smallcaption{\label{figSketchesVWXY}
   A leader always gets prepared,
   as in (w), (x), or (y),
   so that it will be positioned correctly,
   as in (v),
   to read an element of tape data.
   In (y),
   since $c$ is a multiple of 6, the total \ovd\ mod 4
   is $1+0+3+(2c-1)\cdot (0+3)+0=6c+1=1$, as needed in (v).
 }\vspace{-6pt}
\end{figure}

\begin{figure}
 \centerline{
   \setlength{\unitlength}{1pt}\thicklines
   \beginpicture{170}{210}
    \put(0,20){\beginpicture{170}{190}\put(95,-15){\makebox(0,0){\sklab{z}}}
\put(10,120){\makebox(0,0){$A^4$}}
\put(20,110){\line(1,-1){40}}
\put(30,140){\makebox(0,0){$A^4$}}
\put(40,130){\line(1,-1){80}}
\put(103,76){\makebox(0,0){$A$}}
\put(50,160){\makebox(0,0){$A^4$}}
\put(60,150){\line(1,-1){20}}
\put(70,180){\makebox(0,0){$A^4$}}
\put(80,170){\line(1,-1){40}}
\put(111,148){\makebox(0,0){$A$}}
\put(100,180){\makebox(0,0){$C_2$}}
\put(100,170){\line(0,-1){20}}
\put(100,150){\line(-1,-1){20}}
\put(85,145){\makebox(0,0){\ebar}}
\put(80,130){\line(0,-1){40}}
\put(73,115){\makebox(0,0){$C_2$}}
\put(80,90){\line(-1,-1){20}}
\put(65,85){\makebox(0,0){\ebar}}
\put(60,70){\line(0,-1){50}}
\put(60,10){\makebox(0,0){$C_2$}}
\put(120,180){\makebox(0,0){$C_2$}}
\put(120,170){\line(0,-1){120}}
\put(113,100){\makebox(0,0){$C_1$}}
\put(120,50){\line(-1,-4){7.5}}
\put(110,10){\makebox(0,0){$F$}}
\put(140,180){\makebox(0,0){$C_2$}}
\put(140,170){\line(0,-1){150}}
\put(140,10){\makebox(0,0){$C_2$}}
\put(160,180){\makebox(0,0){$C_2$}}
\put(160,170){\line(0,-1){150}}
\put(160,10){\makebox(0,0){$C_2$}}
    \eendpicture}
   \eendpicture
 }
 \smallcaption{\label{figSketchesZ}
   An $F$ is produced if and only if the \tm\ halts.
   The \ts\ in \secref{compiler} is set up so that when the \tm\ halts, no more appendants are
   produced for the tape.  This will cause ossifier $A^4$s to eventually
   hit a character of tape data, which leads to
   the production of an $F$ as shown.
 }
\end{figure}

\section{A Polynomial Time Simulation}\label{directsim}

In~\cite{Neary06}, Neary and Woods solved the geometry problem described here on \pref{geomprob}
by using the following idea, which we present in a form adapted to \ts s.
(Another adaptation to \ts s is given in Neary's thesis~\cite{Neary08}.)

First of all, note that on a circular tape, it is sufficient for the \tm\ head to always move to the right.
For example, the natural approach to simulating a cellular automaton works in this way:
The \tm\ can store the local configuration of the tape in its finite state,
and can read in the old configuration and write out the new configuration in a single pass.
Extra cells can be written onto the tape at the wrap-around point as necessary, allowing it to grow.

So, when the simulated \tm\ head is at a certain position on the tag system tape,
what we would like to do is to find the next position on the tape to the right of the current position.
The tag system can do this not in a single pass, but in a series of log($n$) cycles,
where each cycle consists of four passes.
On the first cycle, it eliminates every second possibility.
On the next cycle, it eliminates every second remaining possibility, and so on.
After log($n$) cycles, the only remaining possibility is the first one: the next position on the tape.

So that the system can know when log($n$) cycles have occurred,
it keeps on its tape a power of 2 that is larger than $n$, called the counter.
It performs this same halving-of-possibilities process on the counter,
where it is able to detect the end of the process since then for the first time there will
be an odd number of possibilities remaining.

When the tape grows in length past a power of two,
then the counter must be doubled in size.
This condition is detected by keeping track of a ``growth flag''
which is set at the beginning of each series of cycles,
and then if the tape is ever noticed to have a non-power-of-two length,
then the flag is reset.
If the flag is still set at the end of the series of cycles,
and the tape is being extended,
then the counter size is doubled.

In \ts s, one must take care that the counter behaves properly
when the \tm\ ``head'' jumps across it, since then the counter
gets processed twice before the \ts\ head is at the beginning of the tape again.

In the \ts , we keep all sorts of information in the symbols,
using a large alphabet, since the communication bandwidth
between symbols is very low (a single bit or less per pass).
For example, every symbol on the tape knows the current
state of the \tm\ and the current stage of the simulation algorithm.
We use six stages, and each stage does one pass over the tape.

We will use the following symbols:
\begin{itemize}
\item
For the head: H, h, P, Q \newline
 P is used for a head remembering ``A'', Q for ``B''.
\item
For the counter: U u X x V v Y y \newline
 X is an eliminated U, and V and Y are a form of U and X that represent the growth flag having been reset.
\item
For the tape: A a B b C c D d \newline
 A and B represent the two values used on the binary \tm\ tape.
 C and D are the eliminated forms of A and B.
\end{itemize}

The lower-case letters are used when we need to be able to detect
which of the two positions is being read or ignored.

If the \tm\ has $k$ states, then there are actually $6k$ symbols for each of the above symbols.
We could write these for example as $H_{4,7}$ being
the $H$ symbol that is used in stage 4 when the \tm\ is in state 7,
but usually we will omit the indices, to avoid clutter.
Not all of the 6k symbols are used, since not all of the letters are used in all stages.

For clarity, we will also use ``-'' as a symbol,
to be used in positions where we know it will not be read,
and we will use ``0'' as a symbol with an empty appendant.
These two symbols do not need any subscripts.

\segsec{Stages}

There is a cycle of four stages, with stage 4 going back to stage 1
until the first tape symbol has been isolated,
at which point stage 4 goes on to stage 5 and then 6.
Stage 6 performs the simulation of a step of the \tm\ and
jumps back to stage 3 to start isolating the next tape symbol.

We will start in stage 2, because it has the simplest form.
At the front is Hh, then the tape is some combination of AA and BB,
and somewhere there is a power of 2 of Uu's, at least as long as the A/B portion of the tape.

We will explain the stages by following an example.

\segsec{Stage 2}

Example:   H h A A U u U u U u U u B B A A    (\tm\ tape is ``A,B,A'')
\newline
Incoming parity:  If reading second symbol of each pair, then the growth flag is unset.
\newline
Outgoing parity:  Start with first symbol of each pair.
\newline
Main change:  Cut number of U's in half, and check whether counter size should be doubled.
\newline
Transfer to stage 3:
\begin{verbatim}
H *   :   H -
h *   :   - H -
A *   :   A A
B *   :   B B
C *   :   C C
D *   :   D D
U *   :   U
u *   :   V
X *   :   X X
x *   :   Y Y
V *   :   V
Y *   :   Y Y
\end{verbatim}

\segsec{Stage 3}

Example:   H - A A U U U U B B A A
\newline
Incoming parity:  Start with first symbol of each pair.
\newline
Outgoing parity:  If reading second symbols, then there was just one U.
\newline
Main change:  Change odd U's (first, third, etc.) to UuXx, erase even U's.
\newline
Transfer to stage 4:
\begin{verbatim}
H *   :   H h
A *   :   A a
B *   :   B b
C *   :   C c
D *   :   D d
U *   :   U u X x
V *   :   V v Y y
X *   :   X x
Y *   :   Y y
\end{verbatim}

\segsec{Stage 4}

Example:   H h A a U u X x U u X x B b A a
\newline
Incoming parity:  If reading small letters, then first tape symbol is isolated, and we will go on to stage 5.  Otherwise we go back to stage 1.
\newline
Outgoing parity:  Start with the first symbol of each pair.
\newline
Main change to stage 1:  Change A to Aa0, and B to Bb0.
\newline
Transfer to stage 1:
\begin{verbatim}
H *   :   H -
A *   :   A a 0
B *   :   B b 0
C *   :   C C
D *   :   D D
U *   :   U U
V *   :   V V
X *   :   X X
Y *   :   Y Y
\end{verbatim}

\segsec{Stage 1}

Example:   H - A a 0 U U X X U U X X B b 0 A a 0
\newline
Incoming parity:  We read the H.
\newline
Outgoing parity:  If there were an even (resp. odd) number of A's and B's, then we will read the first (resp. second) symbol of each pair.
\newline
Main change:  Every second A or B is turned into a C or D.
\newline
Transfer to stage 2:
\begin{verbatim}
H *   :   H h
A *   :   A A
a *   :   C C
B *   :   B B
b *   :   D D
C *   :   C C
D *   :   D D
U *   :   U u
V *   :   V V
X *   :   X x
Y *   :   Y Y
\end{verbatim}

\segsec{Stage 2}
Example:   H h A A U u X x U u X x D D A A     (read h)

\segsec{Stage 3}
Example:   - H - A A V Y Y V Y Y D D A A    (read H)

\segsec{Stage 4}
Example:   H h A a V v Y y Y y Y y D d A a    (read H)

\segsec{Stage 1}
Example:   H - A a 0 V V Y Y Y Y Y Y D D A a 0    (read H)

\segsec{Stage 2}
Example:   H h A A V V Y Y Y Y Y Y D D C C    (read H)

\segsec{Stage 3}
Example:   H - A A V Y Y Y Y Y Y D D C C    (read H)

\segsec{Stage 4}
Example:   H h A a V v Y y Y y Y y Y y D d C c    (read h)
\newline
Main change to stage 5:  H and U disappear.  Isolated A/B becomes P/Q.  C/D turn back into A/B.  X/Y becomes a mixed-stage form UX/VY.  The ``head'' has jumped from the H to the P/Q, so it has jumped onto the symbol it is reading.  This may cause it to jump over the counter (the X's/Y's), which will be discussed below.
\newline
Transfer to stage 5:
\begin{verbatim}
h *   :
a *   :   - P -
b *   :   - Q -
c *   :   A -
d *   :   B -
u *   :
v *   :
x *   :   U x
y *   :   V y
\end{verbatim}

Note regarding the transfer:
The x and y symbols are written with a stage 4 subscript!

\segsec{Stage 5}

Example:   - P - V y V y V y V y B - A -    (read P)
\newline
Incoming parity:  We read the P or Q.
\newline
Outgoing parity:  Read first symbols if P, read second symbols if Q.
\newline
Main change:  x and y ``stage 4'' symbols disappear (more on them below), and parity is set to reflect P vs. Q.
\newline
Transfer to stage 6:
\begin{verbatim}
P *   :   P -
Q *   :   Q
A *   :   A a
B *   :   B b
U *   :   U u
V *   :   V v
\end{verbatim}

\segsec{Stage 6}

Example:   P - V v V v V v V v B b A a    (read P)
\newline
Incoming parity:  If P at front, will read first of each pair.  If Q at front, will read second of each pair.
\newline
Outgoing parity:  Will read first of each pair.
\newline
Main change:  A step of the \tm\ is simulated.  The state (invisible subscript) changes.  The new symbols get written onto the tape.  V's turn back into U's (i.e. the growth flag is reset).  The counter size (number of U's) doubles if necessary.  The counter size is also halved for direct entry to stage 3.
\newline
Transfer to stage 3:
\begin{verbatim}
P *   :   [A a B b] H -
Q *   :   [A a B b] - H -
A *   :   A A      (new state when reading an A)
a *   :   A A      (new state when reading a B)
B *   :   B B      (new state when reading an A)
b *   :   B B      (new state when reading a B)
U *   :   U [U]    (new state when reading an A)
u *   :   U [U]    (new state when reading a B)
V *   :   U        (new state when reading an A)
v *   :   U        (new state when reading a B)
\end{verbatim}

Notes regarding the transfer:
The [A a B b] portions depends on what symbol(s) are being written,
and they are written with subscripts for stage 6,
and the current (old) state (and they will be the last thing processed at the end of this transfer),
while the H (and everything below) gets written with the new state and stage 3.

The [U] portions are included only when the corresponding \tm\ transition writes 2 symbols.
This is where the length of the counter gets doubled.

\segsec{Stage 3} Example:   H - U U U U B B A A B B B B

\segsec{Stage 4} Example:   H h U u X x U u X x B b A a B b B b

\segsec{Stage 1} Example:   H - U U X X U U X X B b 0 A a 0 B b 0 B b 0

\segsec{Stage 2} Example:   H h U u X x U u X x B B C C B B D D

\segsec{Stage 3} Example:   H - U X X U X X B B C C B B D D

\segsec{Stage 4} Example:   H h U u X x X x X x B b C c B b D d

\segsec{Stage 1} Example:   H - U U X X X X X X B b 0 C C B b 0 D D

\segsec{Stage 2} Example:   H h U u X x X x X x B B C C D D D D

\segsec{Stage 3} Example:   H - U X X X X X X B B C C D D D D

\segsec{Stage 4} Example:   H h U u X x X x X x X x B b C c D d D d    (read h)

Now we will see what the mixed-stage stuff is doing, where the stage-4 ``x'' turns into a ``U x'' pair with the U in stage 5 but the x in stage 4.  Recall that h and u both disappear.

\segsec{Stage 4.5} Example:   U x U x U x U x - Q - A - B - B -    (read x)

Now we see that that the stage-5 U's will get ignored as stage 4 finishes by processing the counter a second time.

\segsec{Stage 5} Example:   - Q - A - B - B - U x U x U x U x

And we see that this time the stage-4 x's will get ignored during the stage-5 processing.  Stage 5 doesn't do much because it is mostly serving as a signal to the counter regarding what stage is happening.  The counter knows it is still stage 4 while the second symbols are getting read, and the counter knows stage 5 has arrived when the first symbols get read.  So stage 5 has to have a fixed parity during its processing, making it a boring stage.

\segsec{Stage 6} Example:   Q A a B b B b U u U u U u U u

And so on...  The counter will double in size on the next step if necessary.  It works well in simulation.

As mentioned in~\cite{Woods06}, the fact that tag systems can compute efficiently
means that all known small universal Turing machines work in polynomial time.

\bibliographystyle{eptcs}

\end{document}